\documentclass{amsart}

\usepackage[margin=1in]{geometry}
\usepackage{amsthm}
\usepackage{amsmath}
\usepackage{amssymb}
\usepackage{bbm}
\usepackage{enumitem}
\usepackage{xcolor}
\usepackage{placeins}
\usepackage{tikz}
\usepackage{graphicx}
\usepackage{algorithm}
\usepackage{algpseudocode}
\usepackage{xspace}
\usepackage{hyperref}

\graphicspath{{figures/}{writeup/figures/}}

\usepackage[draft]{todonotes}
\setuptodonotes{
    xshift=1cm,
    size=\scriptsize,
    fancyline=false,
    bordercolor=white
}

\title{Towards stratified sampling for redistricting plans}
\author[ZW, GH, JY, JM, AG]{Zijian Wang, Gregory J. Herschlag, Joon-Hyeok Yim,\\ Jonathan C. Mattingly, and Anna C. Gilbert}
\thanks{%
(ZW) Department of Applied and Computational Mathematics, Yale University, New Haven, CT 06511, USA\\
\indent(GH) Department of Mathematics, Duke University, Durham, NC 27708 USA\\
\indent(JY) Department of Electrical \& Computer Engineering, Yale University, New Haven, CT 06511, USA\\
\indent(JM) Department of Mathematics, Duke University, Durham, NC 27708 USA\\
\indent(AG) Department of Statistics \& Data Science, Yale University, New Haven, CT 06511, USA\\
\indent E-mail addresses: \url{zijian.wang@yale.edu}, \url{gjh@math.duke.edu}, \url{joonhyeokyim@gmail.com}, \url{jonm@duke.edu}, \url{anna.gilbert@yale.edu}\\
\indent Date: \today
}

\newcommand{\be}{\mathbb{E}}

\newcommand{\ind}{\mathbbm{1}}

\newcommand{\hccultra}{\textsc{HccUltraFit}\xspace}

\begin{document}

\begin{abstract}
Rapid algorithmic developments have accelerated the sampling of redistricting ensembles (balanced graph partitions), yet evaluating rare events and sampling complex target measures remains a core challenge due to the high-dimensional and combinatorial nature of the phase space.
We address a prerequisite for stratified sampling on this space: constructing and diagnosing candidate strata with suitable coverage and overlap.
We build a grammar on observed plans by clustering districts into representative ``letters'' and using them to form plan-level ``words.''
A partition of unity over these words gives a soft assignment of plans to strata and allows us to estimate stratum masses and an overlap-induced flux matrix.
We demonstrate this computational pipeline using real-world congressional redistricting data from Connecticut and examine how strata learned from one target distribution behave under related distributions.
The resulting construction provides a foundation for future stratified sampling on spaces of redistricting plans or balanced graph partitions.
We do not implement a complete stratified sampler here; evaluating whether the proposed strata improve sampling efficiency or reduce estimator variance is left for future work.
\end{abstract}

\maketitle
\section{Introduction}
Members of the U.S. House of Representatives are elected every two years, with a total of $435$ seats apportioned among the states based on population counts from the census~\cite{eckman2019apportionment}.
States that have more than one seat must draw congressional districts and elect one representative per district.
As population distributions shift, district boundary lines are usually redrawn after each census, subject to both federal and state rules.
Since the exact placement of these boundaries can heavily influence election outcomes, the redistricting process is persistently controversial.
Disputes over partisan or racial gerrymandering frequently lead to extensive litigation.
To objectively evaluate whether a redistricting plan has been unfairly manipulated, courts increasingly take into account computational methods~\cite{RuchoVCC,LWVvPA,LewisVCommonCause,HarperVHallMoore,AllenVMilligan,deford2019redistricting,duchinPAreport,CovingtonvNC,mattingly2019expert,mattingly2021expert}.

Mathematically, a redistricting plan is a partition of a state's basic geographic or voting units (e.g., precincts) into a fixed number of districts.
These units naturally form a planar graph that encodes geographical contiguity constraints.
A standard way to quantify whether an enacted or proposed redistricting plan is unusual is to compare it against an ensemble of plans that follow similar legal and geographical constraints.
To produce statistically meaningful counterfactuals and hypotheticals, one measures observables of interest~\cite{mattingly2014redistricting,fifield2015new,herschlag2020quantifying,Chen15} like voting patterns, demographic distributions, etc., on the ensemble to see whether the enacted plan is typical with respect to the ensemble.
Such ensembles are typically specified via score functions that encode constraints and preferences.
This inspires a large body of work related to drawing ``representative'' redistricting plans at random (from some policy related probability measure) from all such allowable plans, typically done via MCMC simulations.
Recent algorithmic advances make it feasible to generate large ensembles under increasingly nuanced constraints and policy preferences, for instance, via reversible forest-based merge-split samplers~\cite{autry2020multi,autry2023metropolized,cannon2026spanning}, lifted Cycle Walks~\cite{cyclewalk,mcwhorter2025marked,akitaya2026balanced}, sequential Monte Carlo methods~\cite{mccartan2023sequential,o2026generalized}, and more~\cite{gurnee2021fairmandering,chuang2025multiscale}.

Despite rapid progress in sampling algorithms, evaluating rare events and exploring complex target measures remains a persistent challenge in the study of redistricting ensembles.
For instance, it is well known that existing samplers exhibit low acceptance rates and slow convergence when the target measure deviates from the tree-count measure~\cite{autry2020multi,cyclewalk}.
Stratified sampling is a potential variance-reduction technique for these problems, but applying it first requires a useful stratification of the phase space.
Also known historically in computational chemistry, with some variations, as umbrella sampling~\cite{torrie1977nonphysical}, this approach has become a popular technique for sampling rare events, as well as complex measures~\cite{dinner2020stratification}.
Recent works extend this approach to nonequilibrium sampling via trajectory stratification~\cite{dinner2018trajectory}.
A key prerequisite for stratified sampling is that we need strata that cover the phase space, while maintaining suitable overlaps.
The present paper focuses on this prerequisite---constructing and diagnosing candidate strata---rather than implementing a complete stratified sampler.
In computational chemistry applications, the underlying phase space is usually an interval, or, more generally a subset of $\mathbb{R}^n$, where there are several canonical ways to construct strata with desired features. These are often based on physically inspired order parameters or collective variables.
However, it is not immediately clear how to construct such strata over the space of redistricting plans, which is discrete and combinatorial.
Three features make the stratification of redistricting plans particularly challenging: (i) small local boundary moves can produce distinct plans that are, for practical purposes, ``almost the same'', so any useful notion of similarity must capture essential structure while ignoring inconsequential boundary noise; (ii) plans are unordered tuples of districts, so comparing two plans typically requires solving a district-matching problem, which is expensive at scale; and (iii) clustering plans directly can be computationally heavy and may obscure district-level structure, since variation in one district can be largely independent of variation in another.

In this paper, we build a computational pipeline for constructing and diagnosing candidate strata on the space of redistricting plans. 
We address the aforementioned challenges by first clustering at the district level and then using that structure to build plan-level strata.
This district-level clustering step addresses (i) by grouping together similar districts, and it addresses (iii) by retaining district-level granularity that can be obscured if we directly perform plan-level clustering.
We use a beam-search heuristic to identify, for each plan, a small set of low-distance candidate words when constructing the strata.
As the first step of the pipeline, we collect districts appearing across the ensemble and cluster them with a hierarchical correlation clustering algorithm \hccultra~\cite{DBLP:journals/corr/abs-2409-01010} under a population-weighted symmetric-difference metric, producing a small ``alphabet'' of representative districts (``letters'').
Each plan is then encoded as an unordered tuple of letters (a ``word'').
Because only a small fraction of all possible words correspond to viable plans, we construct a representative word set near the observed samples from the ensemble via a beam search that expands partial words district-by-district while retaining only the best candidates.
Next, we define a partition of unity (POU) that softly assigns each plan to nearby words.
We fix a small set of POU parameters and use a long MCMC run to estimate flux matrices and stationary distributions as diagnostics of stratum mass and overlap.
We illustrate the pipeline on Connecticut congressional redistricting data, where the five-district setting makes the constructed words visually interpretable.
We first analyze the baseline tree-count ensemble, then test whether strata learned from this baseline remain useful for a calibrated family of target measures.
These experiments evaluate the representation, coverage, overlap, and cross-distribution stability of the candidate strata; they do not evaluate a complete stratified sampler.
Implementing such a sampler and testing its efficiency and variance reduction against an unstratified baseline remain future work.

\medskip
\noindent\textbf{Road map}
\begin{itemize}
    \item Section~\ref{sec:chap4 background} provides background on the three ingredients in our pipeline: redistricting ensemble sampling, stratified sampling, and hierarchical clustering.
    \item Section~\ref{sec:chap4 notation_setup} formalizes notation for districts and plans; defines district-level and plan-level distances; introduces letters and words; defines the POU, weights and flux matrix.
    \item Section~\ref{sec:chap4 methodology} details the full pipeline: data generation, hierarchical clustering for building letters, beam search for candidate words, POU diagnostics, and two-stage word pruning.
    \item Section~\ref{sec:chap4 numerical_experiments} applies the pipeline to Connecticut, studies the resulting words and flux diagnostics, and examines how strata learned from $\gamma=0$ behave under calibrated target distributions with $\gamma>0$.
    \item Section~\ref{sec:chap4 future_directions} discusses next steps, including running full stratified sampling using the strata constructed with the pipeline, and updating the alphabet for new data.
    \item Section~\ref{sec:chap4 conclusion} concludes the paper.
\end{itemize}

\section{Background}\label{sec:chap4 background}

\subsection{Redistricting ensemble sampling}\label{subsec:districting_ensembles}

The idea of using a score-based probability distribution to assess redistricting counterfactuals was introduced in $2014$, concurrently by Mattingly and Vaughn~\cite{mattingly2014redistricting,herschlag2020quantifying} and Fifield et al.~\cite{fifield2015new,fifield2020essential}.
Tree-based merge-split samplers have quickly emerged as the standard method to draw samples from ensembles of redistricting plans~\cite{deford2021recombination}.
In the most common variant of the recombination (RECOM) algorithm, one repeatedly merges two adjacent districts, samples a random spanning tree (using Wilson's algorithm via loop-erased random walk), and then cuts an edge to form two new districts that satisfy population balance constraints.
Since each step produces two entirely new districts from their union, RECOM makes relatively global movements compared to local node flips.
This implies strong empirical mixing properties in practical applications.
It is worth noting that the spanning-tree construction induces an implicit bias toward districts with a high number of spanning trees, which has since been correlated with, and even used as, a measure of compactness~\cite{deford2021recombination}.

A key limitation of vanilla RECOM is that its stationary distribution is unknown and may differ from a specified target measure.
Moreover, using RECOM as a Metropolis-Hastings proposal (without modification) is infeasible because the forward and reverse proposal probabilities are computationally intractable.
The difficulty mainly comes from the fact that there are many possible spanning trees that can lead to the newly proposed districts, and each tree has a different probability of being cut to lead to the new districts.
To address this problem, Metropolized Forest RECOM keeps track of spanning forests instead of just partitions, making forward and backward calculations of proposal probabilities viable~\cite{autry2023metropolized}.
Building on this idea, multi-scale merge-split methods further extend the state space to a forest of tree hierarchies, which has improved scaling properties and is useful for handling county constraints~\cite{autry2020multi}.

In contrast to merge-split proposals that completely resample a pair of districts, cycle-walk-like algorithms lift the space to a forest (or tree) and then create and delete cycles: these methods add an edge (or edges) to create a cycle and then remove another edge (or edges) at random to break the cycle~\cite{cyclewalk,mcwhorter2025marked,akitaya2026balanced}.
This produces relatively local moves compared to the methods discussed earlier.
The resulting chain can be Metropolized to target and then efficiently sample from a wider class of user-specified distributions.
Specifically, numerical studies show good convergence performance on examples that are challenging for existing Recom-type algorithms.
In this paper, we use Metropolized Cycle Walk to generate the redistricting plans as input to our computational pipeline~\cite{cyclewalk}.

\subsection{Stratified Sampling}\label{subsec:umbrella_mbar}
Stratified sampling is a localized variant of importance sampling that was developed in computational chemistry to overcome the inefficiencies that can arise in more standard Monte Carlo methods.
Originally introduced in the work by Torrie and Valleau to compute free-energy differences under the name umbrella sampling~\cite{torrie1977nonphysical}, the method and its many extensions~\cite{kumar1992weighted,shirts2008statistically} are increasingly formalized in the modern statistical literature as stratified sampling or stratified MCMC methods~\cite{dinner2018trajectory}. In some variants, a potential is added to localize the dynamics in a region of phase space, in others moves out of localized strata are rejected or killed and replaced by a restarted trajectory. In some methods the strata are simply a partition of the phase space into cells only overlapping on the boundary, while others utilize a soft partition with overlapping cells that generates a partition of unity. While the strata developed here could be used in many algorithms, we will typically discuss the EMUS algorithm using a soft partition.

One core challenge in this type of problem is that a chain rarely crosses large free-energy barriers, making it hard to sample complex measures and rare events~\cite{kastner2011umbrella,kumar1992weighted}.
This challenge closely mirrors the algorithmic bottlenecks when sampling redistricting plans.
It is known that Recom-type algorithms exhibit prohibitively low acceptance rates when the target measure deviates from the tree-count measure~\cite{autry2020multi}.
While newer approaches like Cycle Walk~\cite{cyclewalk,mcwhorter2025marked} show empirical improvement, convergence may still be relatively slow on certain measures of interest.

Stratified sampling addresses this by splitting $\pi$ into easier-to-sample distributions $\{\pi_s\}_{s\in\mathcal{S}}$.
To obtain these distributions, we tilt $\pi$ by nonnegative functions $\{\rho_s\}_{s\in\mathcal{S}}$:
\begin{align}
    \pi_s(x):=\frac{1}{z_s}\rho_s(x)\pi(x),
\end{align}
where $z_s$ is the partition function that normalizes $\pi_s$ into a probability distribution.
The functions $\rho_s$ are designed to cover the phase space, including the low-probability regions, while maintaining overlap to allow communication between different $\pi_s$.
The main challenge for this type of method is to find the normalization terms $z_s$.
Let $z$ be the vector formed by concatenating $z_s$.
The following fact about $z$ can be verified via straightforward matrix multiplication:
\begin{align}\label{eqn: umbrella fixed point}
    zG(z)=z,\quad\quad G_{s,s'}(z):=\mathbb{E}_{\pi_s}\left[\frac{\rho_{s'}/z_{s}}{\sum_{t}\rho_{t}/z_{t}}\right].
\end{align}
The above implies that $z$ is a left eigenvector of $G(z)$ with eigenvalue one.
This provides a fixed-point iteration approach to solve $z$ by alternating between updating $G$ and evaluating its left eigenvector $z$, which is essentially the basis of the eigenvector method for stratified sampling (EMUS)~\cite{dinner2020stratification,matthews2018umbrella}.

Alternatively, one can construct stratified sampling from a soft assignment viewpoint, which forms a more general mathematical framework that easily extends to nonequilibrium sampling~\cite{dinner2018trajectory}.
Suppose we have a set of strata $\mathcal{S}$ equipped with a partition of unity (POU) on the phase space $\{\phi_s\}_{s\in \mathcal{S}}$, which can be viewed as a type of ``soft assignment'' to the strata.
A common way to obtain a POU is to start with functions $\psi_s(x)$ and normalize them:
    \begin{align}
        \phi_s(x):=\frac{\psi_s(x)}{\sum_{s'\in\mathcal{S}}\psi_{s'}(x)},\quad\quad x\in X.
    \end{align}
This computation is tractable as long as each $x$ is only in the support of $\psi_s$ for a small number of $s$.
Using the POU, we can define stratum weights and a flux matrix via
    \begin{align}
        z_s:=\mathbb{E}_\pi[\phi_s],\quad\quad F_{s,s'}:=\mathbb{E}_{\pi_s}[\phi_{s'}].
    \end{align}
Note that expanding $F_{s,s'}$ using $\phi_s$ gives us an equation that is similar to \eqref{eqn: umbrella fixed point}.
However, the flux matrix $F$ constructed in this way is row-stochastic, whereas $G$ is not in general.
One advantage of using this POU approach is that we can view it as the transition kernel of a Markov chain on the set of strata.

\subsection{Hierarchical clustering}\label{subsec:hierarchical_clustering}
Note that the space of redistricting plans is complex, hence it is generally hard to define meaningful strata.
To address this, we analyze the space of individual districts using hierarchical clustering.
Let $\mathcal{X}=\{x_1,x_2,\dots x_N\}$ be a dataset of size $N$, and $D\in\mathbb{R}_{\geq 0}^{N\times N}$ be the pairwise distances.
A hierarchical clustering algorithm maps $D$ to an ultrametric distance matrix $D_T$.
The resulting output of a hierarchical clustering algorithm can be viewed in several mathematically equivalent ways: a nested sequence of partitions on $\mathcal{X}$, a binary tree with $N$ leaves (a dendrogram), or an ultrametric topology on $\mathcal{X}$~\cite{murtagh2012algorithms}. 
In our application, $N$ is the number of distinct districts pooled from the sampled plans based on real-world congressional districting data from Connecticut.
We sampled $50$K plans, each containing $5$ congressional districts, yielding $250$K district occurrences.
Fewer than $0.1\%$ of these were duplicates, which we removed before clustering.

\begin{figure}[!htbp]
    \centering
    \includegraphics[width=0.8\linewidth]{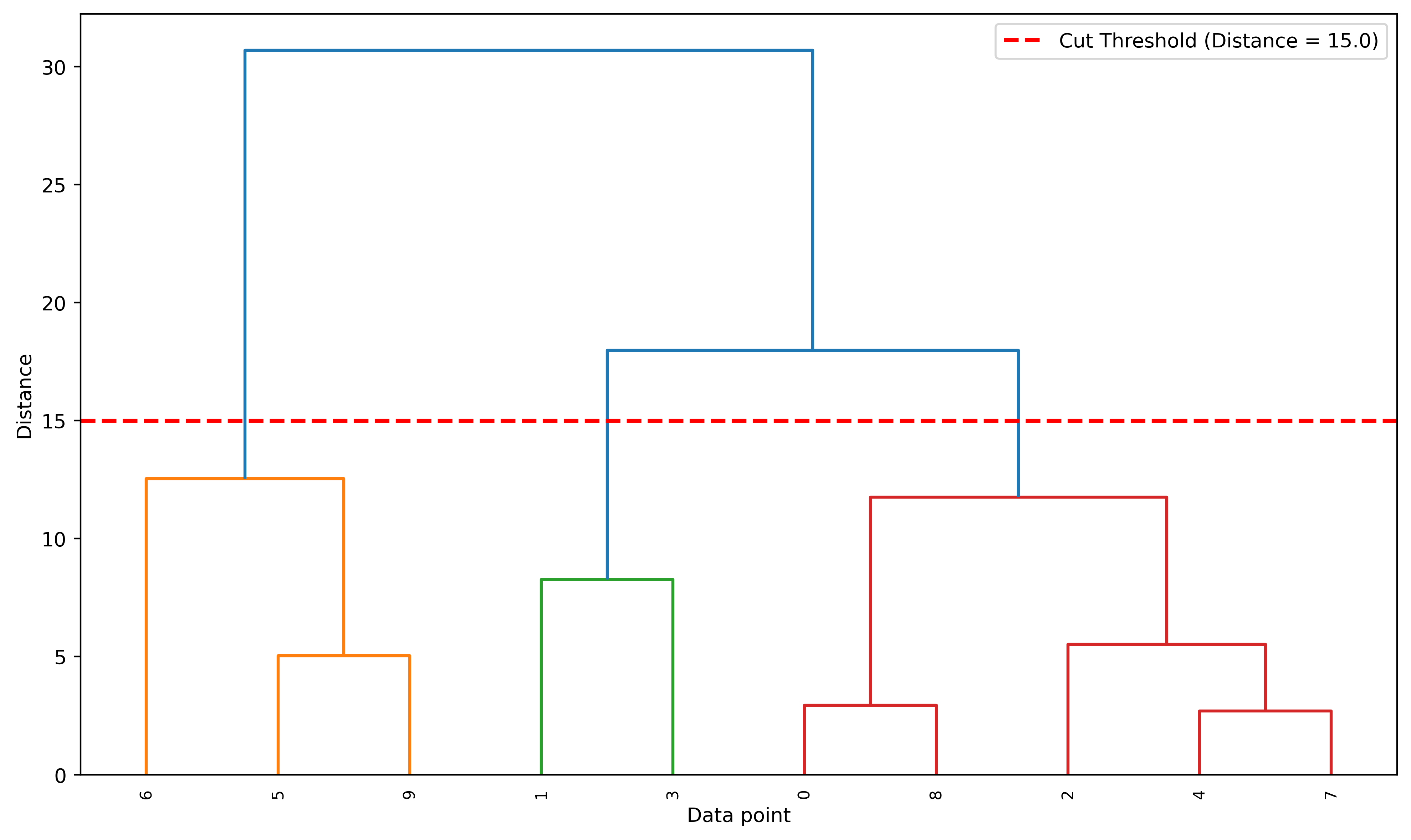}
    \caption{Example dendrogram: $x$-axis shows the indices of data points ($10$ in total), $y$-axis shows the ultrametric distance. The red dashed line indicates a cut at distance $15$, producing three clusters (colored in orange, green and red).}
    \label{fig:dendro_example}
\end{figure}

The nodes of a dendrogram are subsets of $\mathcal{X}$, which can also be viewed as clusters of data points.
In a dendrogram, singleton leaves $\{x_i\}$ gradually merge until they form the root $\mathcal{X}$.
By construction, any two nodes within a dendrogram are either completely disjoint, or one is entirely contained within the other.
The vertical axis of a dendrogram represents the ultrametric distance, a metric that satisfies the strong triangle inequality:
    \begin{align}
        d(x_i,x_j)\leq\max(d(x_i,x_k),d(x_j,x_k)),\quad\quad x_i,x_j,x_k\in\mathcal{X}.
    \end{align}
This condition implies that the distance between any two leaves is simply the height of their lowest common ancestor in the dendrogram.
To extract a clustering of size $K$, one simply ``cuts'' the dendrogram at a specific threshold.
In other words, we discard the edges above the threshold, allowing the sub-trees below the threshold to form a size-$K$ partition of $\mathcal{X}$.
See Figure \ref{fig:dendro_example} for an example of a dendrogram, as well as the cutting procedure.

Classical strategies for building the dendrogram typically involve greedy and irreversible steps that merge the closest pair of clusters at each step based on a single scalar-valued heuristic linkage criterion (e.g., single, complete or average linkage)~\cite{sibson1973slink,defays1977efficient,sokal1958statistical}.
While these linkage criteria are easy to implement, they typically do not capture the underlying geometry of the dataset due to their heuristic nature, i.e., the partitions induced by the dendrogram may poorly reflect the original distance structure.
In contrast, \hccultra~\cite{DBLP:journals/corr/abs-2409-01010} addresses this problem by proposing a hierarchical correlation clustering algorithm with triangle objectives. 
This method fits an ultrametric to $\mathcal{X}$ with a theoretical guarantee on the number of disagreement pairs across every partition level. 
Such consistency across the hierarchy is not observed in common linkage-based methods. 
Although the primary goal of \hccultra in~\cite{DBLP:journals/corr/abs-2409-01010} is to manage ultrametric distances, our computational pipeline specifically utilizes their implementation to generate district clusters. Note that we only use the dendrogram output of \hccultra, but the original paper~\cite{DBLP:journals/corr/abs-2409-01010} has a much wider scope.

\section{Problem setup and notation}\label{sec:chap4 notation_setup}

Let $V:=[P]:=\{1,\dots,P\}$ be the set of basic geographic/voting units in a U.S. state and let $w_1,w_2,\dots,w_P$ be the population in each unit.
The map of the U.S. state induces a graph $G=(V,E)$, where two vertices are connected if they are geographically adjacent.
A congressional plan is a partition of the set of precincts $V$ into $I$ districts, where $I$ is determined by the number of U.S. House seats the state has after apportionment~\cite{eckman2019apportionment}.
At a high level, a plan is admissible if it satisfies a number of constraints.
For instance, it needs to be balanced (up to a certain population deviation threshold) and satisfy contiguity (i.e., be geographically connected).

\begin{figure}[!htbp]
    \centering
    \includegraphics[width=1.0\linewidth]{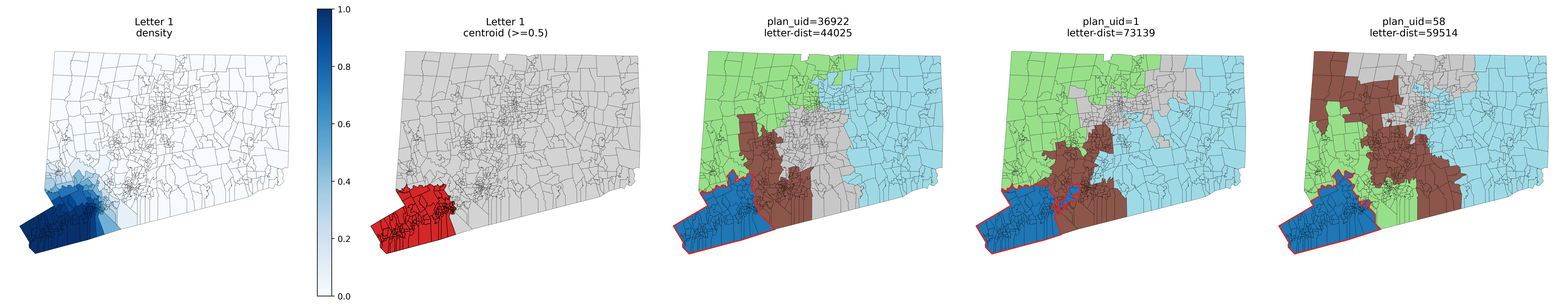}
    \caption{Letter/cluster example based on redistricting samples from Connecticut. From left to right, we have: 1. Fraction of districts containing each precinct; 2. The letter as the centroid of the cluster; 3-5. Plans that contain districts similar to the letter (highlighted in a red outline).}
    \label{fig:letter_district_example}
\end{figure}

Let $\overline{\mathcal V}:=\{0,1\}^P$ be the space of subsets of precincts represented with binary indicators, and let $\mathcal V\subseteq\overline{\mathcal V}$ be the district space.
It is convenient to view a district $v\in\mathcal V$ as a vector, where
\begin{align}
v_p =
    \begin{cases}
    1, & \text{if } p\text{ is in the district},\\
    0, & \text{otherwise.}
    \end{cases}
\end{align}
Let $\overline{\mathcal W}:=\overline{\mathcal V}^{I}/S_I$
be the space of all unordered $I$-tuples of binary precinct indicators,
and let $\mathcal W:=\mathcal V^{I}/S_I$ be the subspace of unordered tuples of districts.
The admissible plan space satisfies
$\mathcal X\subseteq\mathcal W\subseteq\overline{\mathcal W}$.
\footnote{Here $S_I$ is the symmetric group and $\mathcal{W}$ is the quotient space that ensures all elements of $\mathcal{V}^I$ are equivalent if they are permutations of each other.}
A natural way to define a distance $\overline{\mathcal V}\times\overline{\mathcal V}\to\mathbb{R}$ between two districts $v$ and $v'$ is via a capped population-weighted $\ell^1$ distance:
\begin{align}\label{eq:dist_district}
    d(v,v') := \min\Big(d_{\max}, \sum_{p=1}^{P}|v_p -v'_p|\,w_p\Big).
\end{align}
We include a distance upper bound $d_{\max}>0$ to reduce the noise when two districts are disjoint.
A common choice for the cap is 
\begin{align}
d_{\max}=\frac{2}{I}\sum_{p=1}^{P} w_p.
\end{align}
We can extend the distance on $\overline{\mathcal V}$ to a distance function on plans and words in $\overline{\mathcal W}$ by taking the minimum over all matchings:
\begin{align}\label{eq:dist_plan}
    D(w,w')
    := \min_{\pi\in S_I}\sum_{i=1}^I d\big(w_i,w'_{\pi(i)}\big).
\end{align}
One of the central themes in this paper is to systematically identify plans that are similar.
Instead of working with plans directly, we start by identifying similar districts and use that information to reduce the complexity of the plan-level problem. 
In the three redistricting plans shown in Figure \ref{fig:letter_district_example}, we see that the districts in the southwest corner of Connecticut (which includes Fairfield County) are similar to each other, while the rest of the state has drastically different districting patterns.
\FloatBarrier
To identify districts similar to those in Figure \ref{fig:letter_district_example}, we form $K$ clusters on the district space and reduce the complexity of plans by identifying each district by its centroid.
Given a cluster of districts $\widetilde{\mathcal{V}}\subset \mathcal{V}$, we define the centroid $\widetilde v$ to be the point in $\{0,1\}^P$ that minimizes the average uncapped weighted $\ell^1$ distance to all points in $\widetilde{\mathcal V}$.
Under this centroid objective, letters are naturally zero-one valued even if we relax the optimization over $[0,1]^P$.
However, letters are not guaranteed to be an existing district.
An alternative approach is to restrict the distance optimization problem to $\mathcal{V}$.
In this paper, we call these centroids letters.
Using $K$ clusters, we obtain an alphabet of size $K$ and can use these letters to form words, which are unordered tuples of letters of fixed size $I$.
Note that letters are objects in $\{0,1\}^P$ and words are objects in $\overline{\mathcal W}$.

Even after the reduction in district space, we still have an exponentially large word space.
Observe that the space of words relevant to actual plans in the ensemble is much smaller.
For instance, we are not interested in words with repeated letters, assuming that the underlying alphabet forms a good representation of the districts.
Based on this observation, we further reduce the space of words by retaining up to $L$ low-distance candidate words for each observed plan $x\in\mathcal{X}$ under $D$, using the beam-search heuristic with width $L$ described in Section~\ref{sec:chap4 methodology}.
We then work with the union of these candidate sets, which yields a manageable representative word set that is convenient for downstream tasks, e.g. stratified sampling.

Let $\mathcal{S}\subset\overline{\mathcal W}$ be a representative set of words, which forms a set of strata over the space of plans.
We construct a soft assignment of plans to strata using a partition of unity (POU) $\{\phi_s\}_{s\in \mathcal{S}}$ over plans $x\in\mathcal{X}$:
\begin{align}\label{eq:pou}
    \sum_{s\in \mathcal{S}}\phi_s(x)=1,\qquad
    \phi_s(x):=\frac{\psi(x;s)}{\displaystyle\sum_{s'\in \mathcal{S}}\psi(x;s')}.
\end{align}
For each plan $x$, let $\mathcal{N}_{\mathcal{S}}(x)\subseteq\mathcal{S}$ denote the retained word neighborhood for that plan, typically the words in $\mathcal{S}$ that appear among its beam-search candidates.
Note that $\mathcal{N}_{\mathcal{S}}(x)$ is often smaller than the beam-width due to pruning.
We define $\psi$ by
\begin{align}\label{eq:psi_relative}
    \psi(x;s)=
    \begin{cases}
        \exp\!\Big(-T\kappa\big(D(x,s)\big)\Big), & s\in\mathcal{N}_{\mathcal{S}}(x),\\
        0, & \text{otherwise,}
    \end{cases}
\end{align}
where $T\geq 0$ is a temperature parameter controlling sensitivity to distances and $\kappa$ is an $\mathbb{R}\to\mathbb{R}$ function that maps distances (in units of population) into appropriate ranges.
Let $W:=\sum_{p=1}^{P} w_p$ denote the total population. Throughout our numerical experiments, we use the population normalization 
\[
    \kappa(r):=\frac{r}{W},
\]
so that the kernel weights become $\exp\!\left(-T D(x,s)/W\right)$.
When $T=0$, the function $\psi$ assigns equal unnormalized weights to the retained words of a given plan $x$.
Let $\pi$ be the target distribution on plans.
We define the stratum weights by
\[
    z_s := \mathbb{E}_{\pi}[\phi_s],
\]
and the flux matrix between strata by
\begin{align}\label{eq:flux}
    F_{s,s'} := \mathbb{E}_{\pi_s}[\phi_{s'}]
    = \frac{1}{z_s}\mathbb{E}_{\pi}[\phi_s\phi_{s'}],
    \qquad \text{where } \frac{d\pi_s}{d\pi}=\frac{1}{z_s}\phi_s.
\end{align}
By construction, $F$ is row-stochastic and $z$ is stationary for $F$ (i.e., a left eigenvector with eigenvalue $1$).
Observe that this setup is consistent with the standard EMUS stratified sampling setting, see, e.g. \eqref{eqn: umbrella fixed point}.

We conclude this section with a table of notation.
\begin{table}[!htbp]
    \centering
    \small
\renewcommand{\arraystretch}{1.1}
\begin{tabular}{ll}
    \hline
    Symbol & Meaning \\
    \hline
    $G=(V,E)$ & Precinct adjacency graph \\
    $P$ & Number of precincts ($|V|$) \\
    $V$ & Precinct vertex set ($V=[P]$) \\
    $E$ & Adjacency edge set \\
    $\mathcal{V}$ & District indicator space \\
    $\mathcal{W}$ & Space of unordered $I$-tuples of $\mathcal{V}$ \\
    $\mathcal{X}$ & Admissible plan space \\
    $x$ & A plan in $\mathcal{X}$ \\
    $w_p$ & Population weight of precinct $p$ \\
    $W$ & Total population $\sum_{p=1}^P w_p$ \\
    $I$ & Number of districts in a plan \\
    $d_{\max}$ & Cap used in $d$ \\
    $d$ & District/letter distance \eqref{eq:dist_district} \\
$D$ & Plan/word distance \eqref{eq:dist_plan} \\
    $K$ & Number of letters (district clusters) \\
    $l$ & A letter (cluster centroid) \\
    $L$ & Word degree (number of candidates kept) \\
    $\mathcal{S}$ & Representative word set / strata \\
    $s$ & A stratum (word) in $\mathcal{S}$ \\
    $w_i(x)$ & $i$-th retained candidate word for plan $x$ \\
    $\mathcal{N}_{\mathcal{S}}(x)$ & Retained word neighborhood of plan $x$ \\
    $\psi(x;s)$ & Unnormalized kernel \eqref{eq:psi_relative} \\
    $\phi_s(x)$ & POU weight \eqref{eq:pou} \\
    $T$ & Temperature parameter in \eqref{eq:psi_relative} \\
    $\pi$ & Target distribution on plans \\
    $J$ & Target energy, with $\pi(x)\propto e^{-J(x)}$ \\
    $\pi_s$ & POU stratum distribution \eqref{eq:flux} \\
    $z_s$ & POU stratum weight $\be_{\pi}[\phi_s]$ \\
    $F_{s,s'}$ & Flux matrix \eqref{eq:flux} \\
    $q_s$ & Auxiliary biased distribution proposed for future sampling \\
    $J_s^{\mathrm{bias}}$ & Biased energy $J-\phi_s$ defining $q_s$ \\
    $Z_s$ & Normalizer $\be_{\pi}[e^{\phi_s}]$ of $q_s$ \\
    $\overline{\mathcal V}$ & Binary indicator space $\{0,1\}^P$ \\
    $\overline{\mathcal W}$ & Unordered $I$-tuples from $\overline{\mathcal V}$ \\
    \hline
\end{tabular}
\caption{Summary of basic notation.}
\label{tab:notation}
\end{table}

\section{Methodology}\label{sec:chap4 methodology}

\subsection{Data generation}
We study ensembles of redistricting plans $x$ distributed according to a measure of interest:
    \begin{align}
        \pi(x)\propto e^{-J(x)},
    \end{align}
where $J$ is a score function that reflects the hard and soft constraints for a desirable redistricting plan, e.g., population deviation, connectedness, etc.
To efficiently obtain plan samples from $\pi$, we use the Metropolized Cycle Walk~\cite{cyclewalk}.
Each step modifies a spanning forest (on $G$) by adding an edge $e$ and removing a random edge on the cycle created by $e$.
We then accept or reject the resulting forest using a standard Metropolis-Hastings acceptance rule that satisfies detailed balance.
After burn-in, we collect $M$ observations $\mathcal{X}_{obs}:=\{x^{(m)}\}_{m=1}^M$ from the chain output every $100$ steps.

Recall that each plan $x^{(m)}$ is an unordered tuple of $I$ districts in $\mathcal{V}$. 
Pooling districts across the $M$ plans yields a set of $N$ districts
\begin{align}
    \mathcal{D} := \{v_1,\dots,v_N\}\subset \mathcal{V},
\end{align}
where $N\leq MI$ because plans may contain the exact same districts.
We compute the pairwise district-to-district distances according to \eqref{eq:dist_district} and store only distances below $d_{\max}$ in sparse format, avoiding construction of the full dense distance matrix.

\subsection{Building letters}
To find representative districts (``letters''), we cluster the pooled district set $\mathcal{D}$ with \hccultra~\cite{DBLP:journals/corr/abs-2409-01010}, a hierarchical correlation clustering algorithm with theoretical guarantees.
To scale the algorithm to large graphs, we optimize memory usage by tracking only the active roots during merging, which avoids storing a dense $2N\times 2N$ matrix as in a naive implementation.
We bucket-sort the distances and parallelize across bins, where bin boundaries are estimated from a small pilot sample. 
This produces a single dendrogram, which we can cut to obtain a clustering for any desired $K$.

To choose $K$, we use the elbow method on a dispersion objective $\mathcal{L}(K)$.
Let $\mathcal{D}_1,\mathcal{D}_2,\dots,\mathcal{D}_K$ be the clusters, i.e., a partition of $\mathcal{D}$.
For each cluster, we define its letter (centroid) $l_k\in\{0,1\}^P$ as the point that minimizes the sum of distances to members of $\mathcal{D}_k$:
\begin{align}\label{eq:centroid definition}
    l_k &\in \mathop{\mathrm{argmin}}_{l\in\{0,1\}^P}
    \sum_{v\in\mathcal{D}_k}\sum_{p=1}^{P} w_p |l_p-v_p|.
\end{align}
This reduces to a precinct-wise median:
\begin{align}
    (l_k)_p &=
    \begin{cases}
        1, & \displaystyle\sum_{v\in\mathcal{D}_k} v_p \ge \frac{|\mathcal{D}_k|}{2},\\
        0, & \text{otherwise,}
    \end{cases}
    \qquad p=1,\dots,P.
\end{align}
Given the letters $l_1,\dots,l_K$, we score the clustering via the within-cluster dispersion
\begin{align}\label{eq:dispersion_obj}
    \mathcal{L}(K) &:= \sum_{k=1}^K
    \sum_{v\in\mathcal{D}_k}\sum_{p=1}^{P}
    w_p |v_p-(l_k)_p|.
\end{align}
an $\ell^1$ analogue of within-cluster sum of squares (WCSS), which is a common metric for choosing the number of clusters.
Empirically, we find that $\mathcal{L}(K)$ is strongly correlated with standard alternatives such as the Silhouette score, the Calinski--Harabasz index, and BCSS, as well as information-theoretic objectives like BIC and AIC.

\subsection{From letters to words}\label{subsec:letters_to_words}
After clustering districts into an alphabet $\{l_1,\dots,l_K\}$, we construct the set of words by taking the union of unordered $I$-tuples of letters around each plan $x\in\mathcal{X}$.
These candidate words serve two roles: they provide a compressed description of the plans, and they define the local support of the kernel $\psi$ in \eqref{eq:psi_relative} used for the partition of unity.
For a fixed pair of plans, the plan-level distance \eqref{eq:dist_plan} is a linear assignment problem on the $I\times I$ matrix of district-level distances and can be evaluated in $\mathcal{O}(I^3)$ time.
The combinatorial bottleneck instead comes from searching the $\binom{K}{I}$ possible unordered words with distinct letters.
We therefore use a beam-search heuristic to identify up to $L$ low-distance candidate words without enumerating the full word space.

Let $v_1(x),\dots,v_I(x)$ be the districts of $x$.
For each district, we precompute a sorted list of letter candidates
\begin{align}\label{eq:district_to_letter_candidates}
    C_i(x) &:= \big(d_{v_i(x),l_k},l_k\big)_{k=1}^K,\qquad i=1,\dots,I,
\end{align}
where the pairs are sorted by increasing $d_{v_i(x),l_k}$.
During beam search, we incrementally build partial sub-words by adding one letter at a time and keep only the $L$ lowest-cost partial words at each step.
Each partial word is stored as a canonical set of letters; when multiple district-to-letter assignments produce the same set, we retain only the assignment with the smallest accumulated cost.
Because the beam is truncated at intermediate stages, the resulting words are not guaranteed to be the globally closest words under $D$.
After the final stage, we recompute the distance $D(x,s)$ of every retained candidate $s$ exactly as a linear assignment problem.
Including the construction and sorting of the candidate lists, the beam search, and this final exact rescoring, the worst-case running time is
\[
    \mathcal{O}\!\left(IK\log K+ILK\log L+LI^3\right),
\]
assuming constant-time set membership and canonical-key operations.
See Algorithm~\ref{alg:beam_search} for more details.
\begin{algorithm}[!htbp]
\caption{Beam-search heuristic for low-distance candidate words with width $L$}
\label{alg:beam_search}
\begin{algorithmic}[1]
\Require For each district $i\in[I]$, a list $C_i=\{(d_{i,k}, \ell_{i,k})\}_{k=1}^K$ of candidate letters sorted by $d_{i,k}$
\State $B \gets [(0, [])]$ \Comment{beam of (distance, partial word)}
\For{$i=1$ \textbf{to} $I$}
    \State $A \gets$ empty map from canonical partial words to distances
    \ForAll{$(D,s)\in B$}
        \ForAll{$(d,\ell)\in C_i$}
            \If{$\ell \in s$}
                \State \textbf{continue} \Comment{avoid repeated letters}
            \EndIf
            \State $D' \gets D + d$
            \State $s' \gets \Call{CanonicalSet}{s\cup\{\ell\}}$
            \If{$s'\notin A$ \textbf{or} $D'<A[s']$}
                \State $A[s'] \gets D'$ \Comment{keep the best assignment for this partial word}
            \EndIf
        \EndFor
    \EndFor
    \State $H \gets$ empty max-heap keyed by distance
    \ForAll{$(s',D')\in A$}
        \State \Call{HeapPush}{$H,(D',s')$}
        \If{$\Call{HeapSize}{H} > L$}
            \State \Call{HeapPopMax}{$H$} \Comment{discard the largest-distance candidate}
        \EndIf
    \EndFor
    \State $B \gets$ elements of $H$ sorted by increasing distance
\EndFor
\ForAll{$(D,s)\in B$}
    \State $D \gets \Call{AssignmentCost}{s,\{C_i\}_{i=1}^I}$ \Comment{recompute $D(x,s)$ exactly}
\EndFor
\State \Return $B$ sorted by increasing distance
\end{algorithmic}
\end{algorithm}

\begin{figure}[!htbp]
    \centering
    \includegraphics[width=0.55\linewidth]{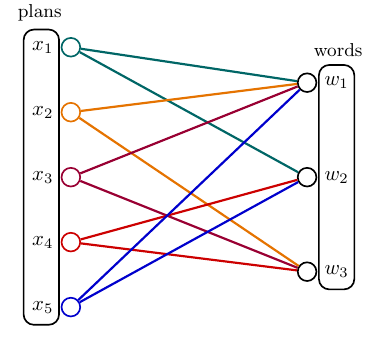}
    \caption{Connectivity of a beam search with width $2$. Edges are colored by plans. Each observed plan on the left connects to the two candidate words retained by the heuristic. In this example, we obtain three words in total.}
    \label{fig:overlap/bipartite}
\end{figure}
\FloatBarrier
For each observed plan $x$, we keep up to $L$ retained candidates as $w_1(x),\dots,w_L(x)$ and take the union of these sets across the ensemble to form the representative word set $\mathcal{S}$ used for stratification.
It may be helpful to treat plans and words as vertices and think of the beam search process as building the connections in a bipartite graph, as shown in Figure \ref{fig:overlap/bipartite}.

\subsection{Towards stratification}
The representative word set $\mathcal{S}$ and the plan-level distances $D$ allow us to construct a stratification of the target distribution $\pi$ on plans.
We start with a partition of unity (POU) $\{\phi_s\}_{s\in\mathcal{S}}$ and define the corresponding stratum distributions $\{\pi_s\}_{s\in\mathcal{S}}$ via tilting.
Each $\phi_s(x)$ can be interpreted as a soft-assignment of plan $x$ to stratum $s$, with flow between strata encoded by the flux matrix $F$ in \eqref{eq:flux}.

One key requirement for stratified sampling to work is coverage, meaning that every plan should have at least one nearby word so that the POU is well-supported.
Before pruning, this is ensured by beam search, which yields up to $L$ candidate words $w_1(x),\dots,w_L(x)$ for each plan $x$ and restricts $\psi(x;s)$ in \eqref{eq:psi_relative} to this neighborhood.
Another important requirement of stratified sampling with soft strata is that strata have appropriate overlaps.
Some overlap is necessary for communication between strata, but excessively large overlap can lead to slow convergence (as well as numerical instability) of EMUS-style fixed-point iterations.
In practice, the balance between overlap and coverage can be controlled by beam-width and pruning threshold, while the temperature $T$ and the choice of $\kappa$ in \eqref{eq:psi_relative} tune the effective overlap within the word neighborhoods.
By construction, the induced coarse dynamics on $\mathcal{S}$ has stationary distribution $z$:
\begin{align}
    \pi(x) &= \sum_{s\in\mathcal{S}} z_s\,\pi_s(x),\qquad
    z^\top F = z^\top,
\end{align}
which we use as a diagnostic to see whether the strata design is suitable for stratified sampling.

Actually running stratified sampling requires additional sampling work (e.g., parallel chains targeting $\{\pi_s\}$), whereas here we focus on setting up a reasonable stratification of plan space and the associated overlap statistics.
When overlap is very large and $\mathcal{S}$ becomes highly redundant, we may prune the strata by solving a set-cover problem induced by the plan--word bipartite graph.
Concretely, each word covers a subset of the observed plans, so the words define a family of subsets of the plan space.
The basic greedy subcover routine takes a boolean adjacency matrix $A\in\{0,1\}^{M\times|\mathcal{S}|}$ and repeatedly adds the word that covers the largest number of currently uncovered plans.
Selecting a minimum-size sub-collection $\mathcal{S}_0\subseteq\mathcal{S}$ can be written as
\begin{align}
    \min_{\mathcal{S}_0\subseteq\mathcal{S}} \quad & |\mathcal{S}_0|,\\
    \text{s.t.}\quad & \sum_{s\in\mathcal{S}_0} A_{m,s} \ge 1,\qquad m=1,\dots,M.
\end{align}
The problem of deciding whether there is a subcover smaller than some fixed size is NP-complete in general, so we use the greedy approximation in Algorithm~\ref{alg:greedy_subcover}.
\begin{algorithm}[!htbp]
\caption{Greedy subcover}
\label{alg:greedy_subcover}
\begin{algorithmic}[1]
\Require Adjacency matrix $A\in\{0,1\}^{M\times|\mathcal{S}|}$
\State $U \gets \{1,\dots,M\}$ \Comment{uncovered plan indices}
\State $R \gets \mathcal{S}$ \Comment{remaining candidate words}
\State $\mathcal{S}_0 \gets \emptyset$
\While{$U\neq\emptyset$ and $R\neq\emptyset$}
    \ForAll{$s\in R$}
        \State $g(s) \gets \sum_{m\in U} A_{m,s}$ \Comment{number of newly covered plans}
    \EndFor
    \If{$\max_{s\in R} g(s)=0$}
        \State \textbf{break}
    \EndIf
    \State pick any $s^\star \in\arg\max_{s\in R} g(s)$
    \State $\mathcal{S}_0 \gets \mathcal{S}_0 \cup \{s^\star\}$
    \State $U \gets \{m\in U: A_{m,s^\star}=0\}$ \Comment{remove newly covered plans}
    \State $R \gets R\setminus\{s^\star\}$
\EndWhile
\State \Return $(\mathcal{S}_0,U)$ \Comment{$U=\emptyset$ when $A$ is fully covered}
\end{algorithmic}
\end{algorithm}

In our setting, the adjacency matrix is obtained by thresholding plan--word distances, so the choice of radius matters.
A large radius improves coverage but may connect plans to distant words, while a small radius gives tighter strata but may leave some observed plans uncovered.
Let $D$ be the plan-to-word distance matrix.
We set the bigger threshold to the worst nearest-word distance over all plans,
\begin{align}
    \tau:=\max_m\min_{s\in\mathcal{S}}D_{m,s},
\end{align}
so every observed plan is covered at radius $\tau$.
For the first stage, let $W:=\sum_{p=1}^P w_p$ be the total population and choose a population-scaled core radius
\begin{align}
    \tau_\eta:=\eta W,\qquad \eta>0.
\end{align}
Thus, $\eta$ is the only parameter that needs to be chosen.
The two thresholds define adjacency matrices
\begin{align}
    A_{\tau_\eta}:= \ind\{D\leq \tau_\eta\},\quad\quad A_\tau:= \ind\{D\leq \tau\},
\end{align}
where the indicator function $\ind$ is applied element-wise.
The first greedy subcover at radius $\tau_\eta$ gives $\mathcal{S}_{\tau_\eta}$.
If any plans remain uncovered, the second greedy subcover at radius $\tau$ is restricted to those plans and to words not selected in the first stage, and gives $\mathcal{S}_\tau$.
We define the pruned set as
\begin{align}
    \mathcal{S}_\eta:=\mathcal{S}_{\tau_\eta}\cup\mathcal{S}_\tau.
\end{align}
Algorithm~\ref{alg:eta_greedy_subcover} summarizes this two-stage procedure.
\begin{algorithm}[!htbp]
\caption{Two-stage greedy subcover with parameter $\eta$}
\label{alg:eta_greedy_subcover}
\begin{algorithmic}[1]
\Require Distance matrix $D\in\mathbb{R}_{\geq 0}^{M\times|\mathcal{S}|}$; total population $W$; threshold parameter $\eta>0$
\State $\tau \gets\max_m \min_{s\in\mathcal{S}}D_{m,s}$
\State $\tau_\eta \gets \eta W$
\State $A_{\tau_\eta}\gets\ind\{D\leq\tau_\eta\}$
\State $A_\tau\gets\ind\{D\leq\tau\}$
\State $(\mathcal{S}_{\tau_\eta},U_{\tau_\eta}) \gets \Call{GreedySubcover}{A_{\tau_\eta}}$
\State $A_\tau^{\mathrm{rem}}\gets A_\tau$ restricted to rows $U_{\tau_\eta}$ and columns $\mathcal{S}\setminus\mathcal{S}_{\tau_\eta}$
\State $(\mathcal{S}_\tau,U_\tau) \gets \Call{GreedySubcover}{A_\tau^{\mathrm{rem}}}$
\State $\mathcal{S}_\eta \gets \mathcal{S}_{\tau_\eta}\cup\mathcal{S}_\tau$
\State \Return $\mathcal{S}_\eta$
\end{algorithmic}
\end{algorithm}

\begin{figure}[!htbp]
    \centering
    \includegraphics[width=0.92\linewidth]{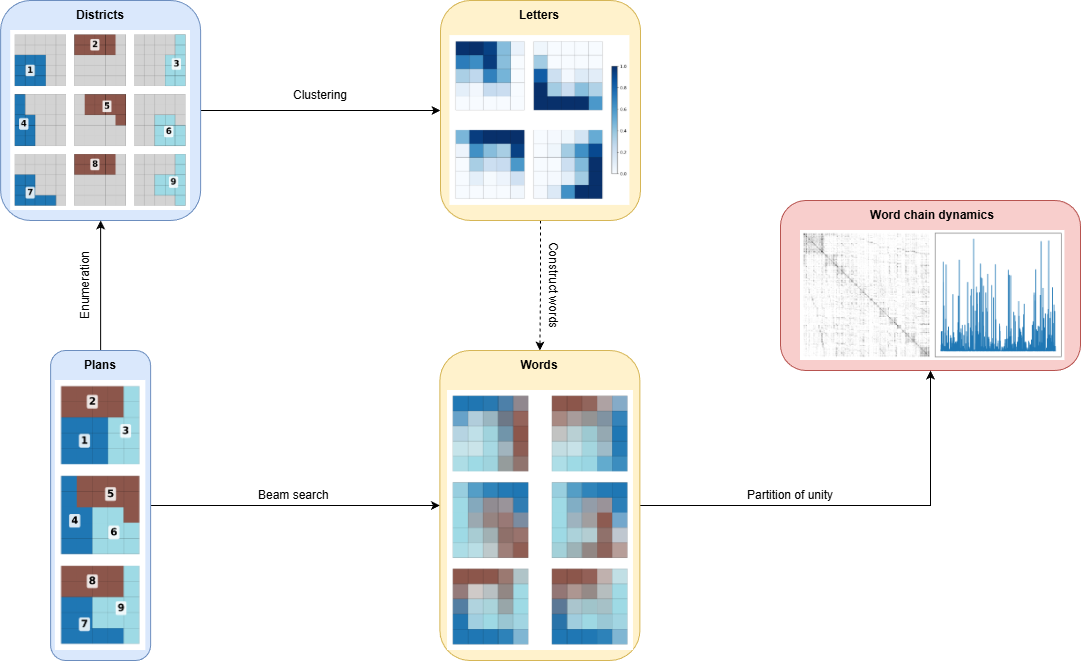}
    \caption{Overview of the pipeline: districts pooled from an ensemble are clustered into representative district types (letters), each plan is encoded as an unordered tuple of letters (a word), and a beam-search heuristic returns a small neighborhood of candidate words around each plan. A partition of unity over the representative word set then yields stratum weights and a flux matrix summarizing coarse dynamics on words.}
    \label{fig:overlap/pipeline_flowchart}
\end{figure}

We end this section by summarizing the methods we use in our analysis pipeline as a flowchart in Figure \ref{fig:overlap/pipeline_flowchart}. We begin with a collection of plans (lower left) which we split into the unique districts (upper left); we cluster the districts into letters (upper middle). For any given plan, we construct a collection of low-distance candidate words from the letters (lower middle). The collection of words gives the strata (after pruning). The overlaps can be used to compute fluxes between strata and the invariant measure of the reduced Markov chain on words.
\FloatBarrier

\section{Numerical Experiments}\label{sec:chap4 numerical_experiments}
In this section, we demonstrate our pipeline using a small-scale real-world example on Connecticut with five districts $|V|=462,|I|=5$.
The data we use is obtained from the VEST project~\cite{mcdonald2008united} via~\cite{redistrictingdatahub}.
\subsection{Exploratory analysis}\label{sec:connecticut}
\begin{figure}[!htbp]
    \centering
    \includegraphics[width=1.0\linewidth]{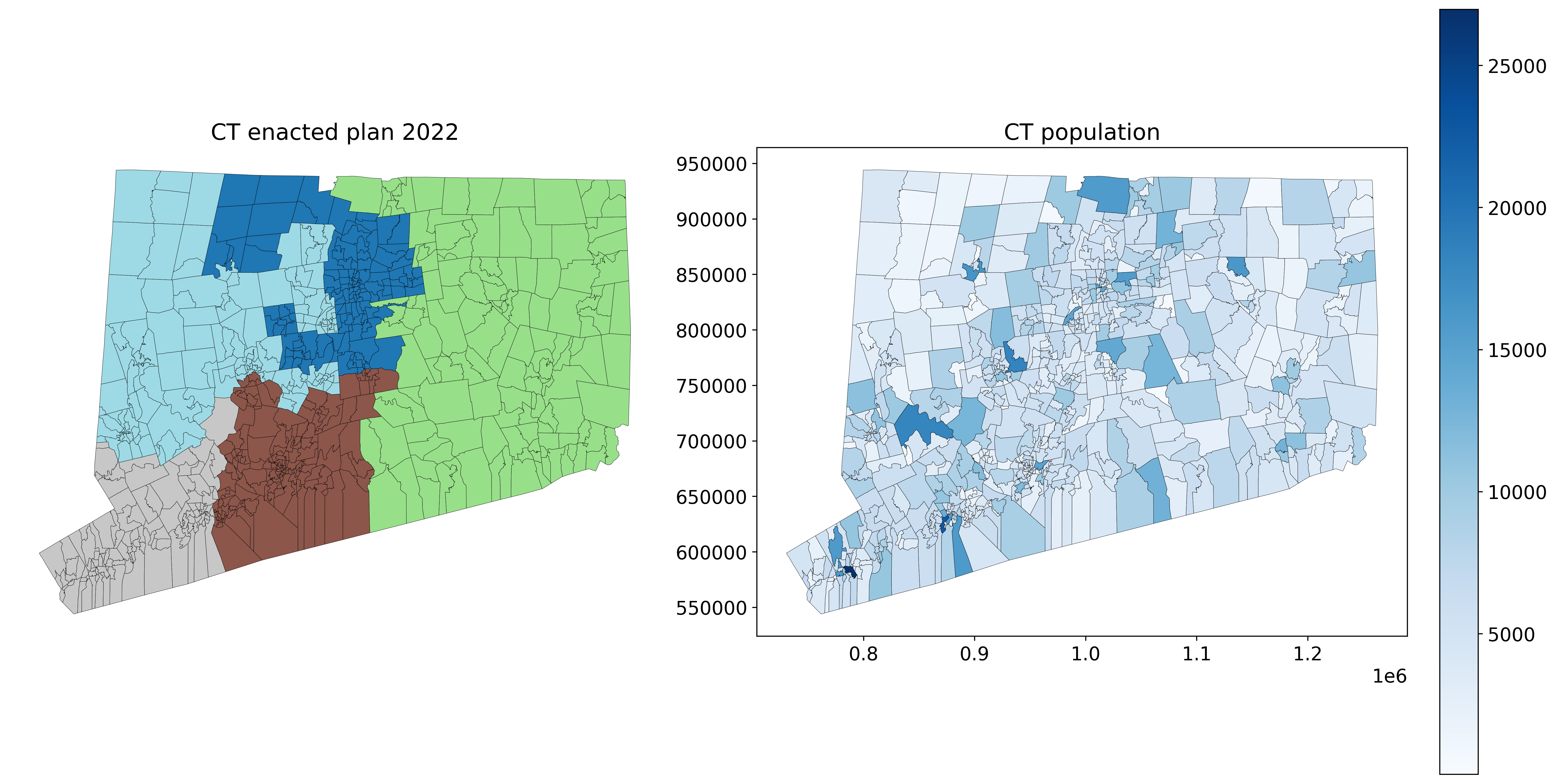}
    \caption{Population distribution for the enacted Connecticut plan.}
    \label{fig:ct_enacted_pop}
\end{figure}

\begin{figure}[!htbp]
    \centering
    \includegraphics[width=1.0\linewidth]{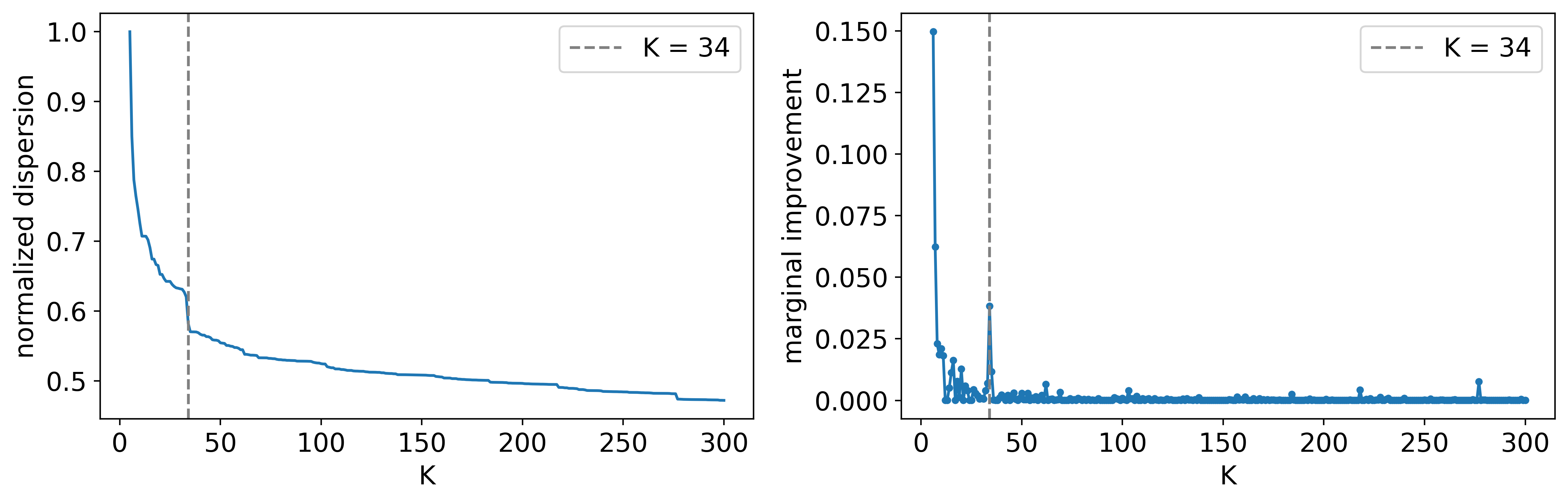}
    \caption{Elbow plots for CT: left, dispersion metric against number of clusters; right, marginal improvements in the dispersion metric by taking consecutive differences. Dispersion shown is normalized by the first value when $K=5$.}
    \label{fig:ct_elbow}
\end{figure}

\begin{figure}[!htbp]
    \centering
    \includegraphics[width=1.0\linewidth]{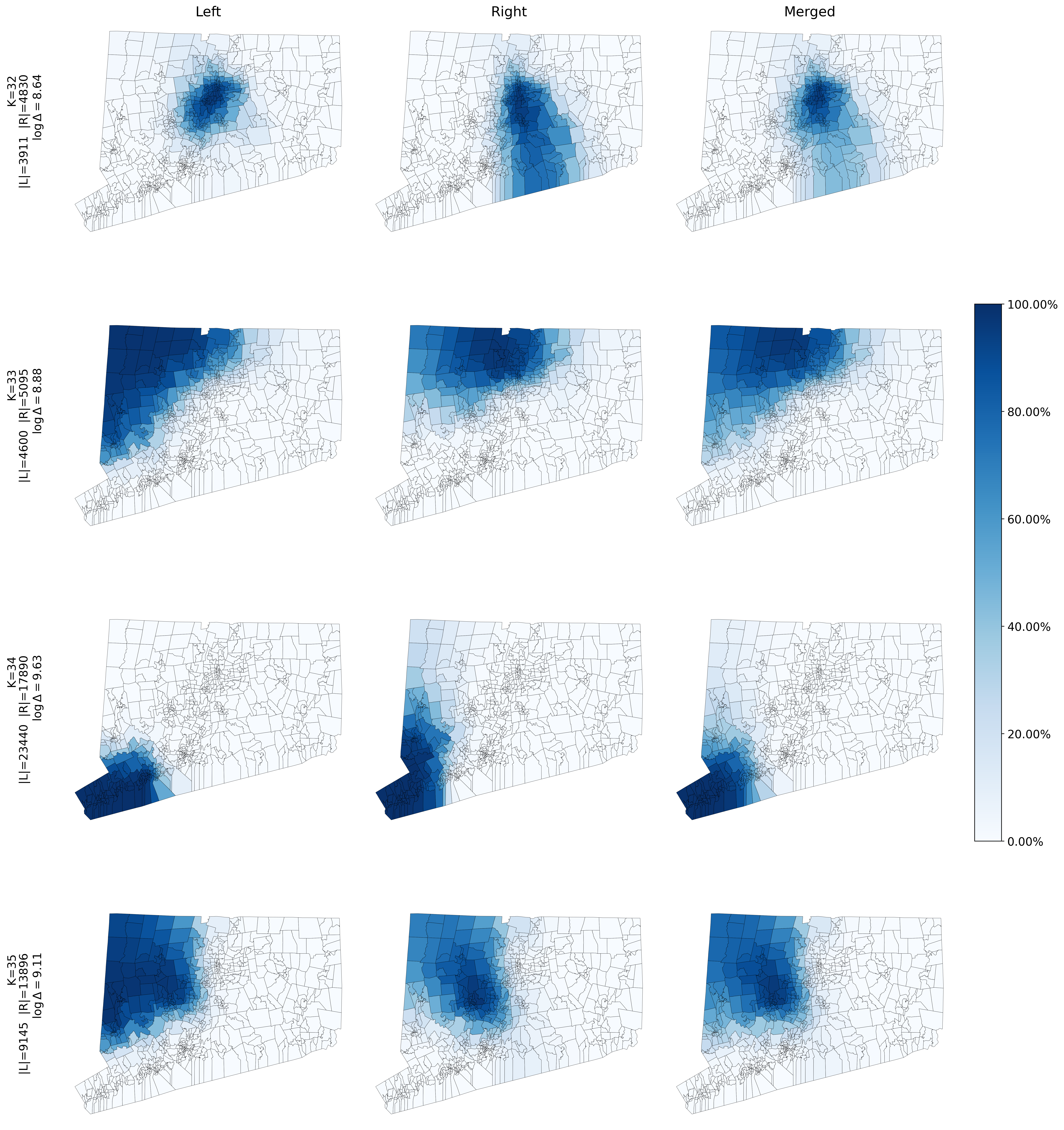}
    \caption{Merges of \hccultra around the elbow. Each row shows two child clusters and their merged cluster for a nearby value of $K$. We annotate each merge by the sizes of the clusters involved, as well as the improvement in the dispersion objective. The color on a precinct $p$ indicates the fraction of districts within a cluster that contains $p$.}
    \label{fig:ct_merge_34}
\end{figure}

\begin{figure}[!htbp]
    \centering
    \includegraphics[width=1.0\linewidth]{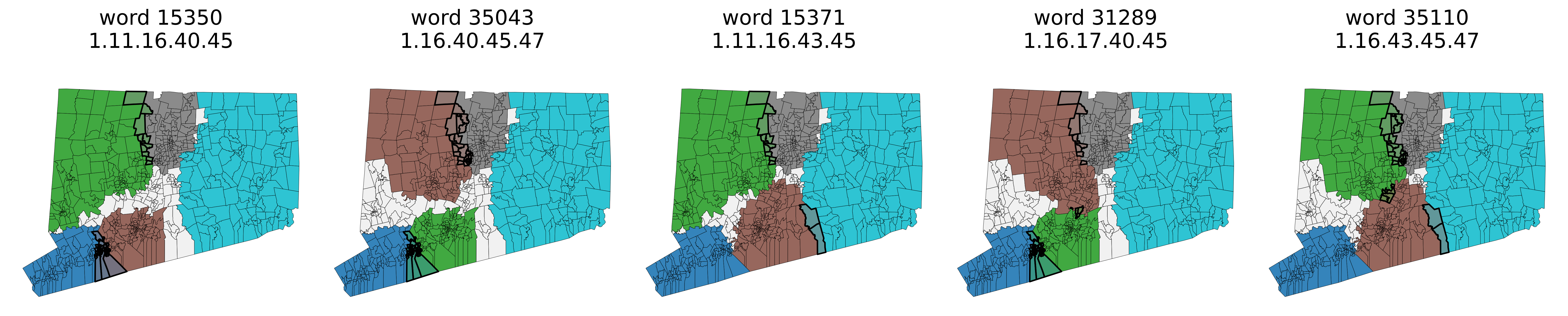}
    \caption{Top five words (by number of plans covered) before pruning. We visualize words by plotting the letters with different colors. Intersections between letters are highlighted in black lines. We display the letter indices below the word, e.g. word 15350 is made up of letters 1, 11, 16, 40, and 45.}
    \label{fig:ct_top5_word_before_prune}
\end{figure}

We apply the pipeline to redistricting plans of Connecticut, where $|V|=462$ precincts are partitioned into $I=5$ congressional districts.
See Figure \ref{fig:ct_enacted_pop} for the enacted plan in $2022$ and its population distribution.
We work with a $50$K-sample Metropolized Cycle Walk run with $\gamma=0$ (i.e., the tree-count measure).
For the approximately $250{,}000$ distinct Connecticut districts, the merging phase of \hccultra took approximately $10$ hours using a single CPU core on the \texttt{bigmem} partition of Yale's Bouchet cluster, whose nodes use Intel Xeon Platinum 8562Y+ processors and $1{,}024$ GB of memory.
We notice a clear ``elbow'' in the CT case, where dispersion drops sharply around $K=34$ as shown in Figure \ref{fig:ct_elbow}.
Figure \ref{fig:ct_merge_34} shows that the merge of two southwestern clusters corresponds to the sharp drop in dispersion, suggesting an optimal alphabet size of $K=34$.
For the word-level analyses below, we use a slightly finer alphabet with $K=50$ and choose a large beam search width of $L=2000$ to capture the relevant words.
Constructing the candidate words by beam search took approximately $1$ hour using a single CPU core on the same Bouchet \texttt{bigmem} partition, with approximately $128$ GB of memory requested.
Figure~\ref{fig:ct_top5_word_before_prune} visualizes the five words with the largest observed plan coverage before pruning; several of these high-coverage words are near-duplicates that differ only by small local letter substitutions.

\begin{figure}[!htbp]
    \centering
    \includegraphics[width=1.0\linewidth]{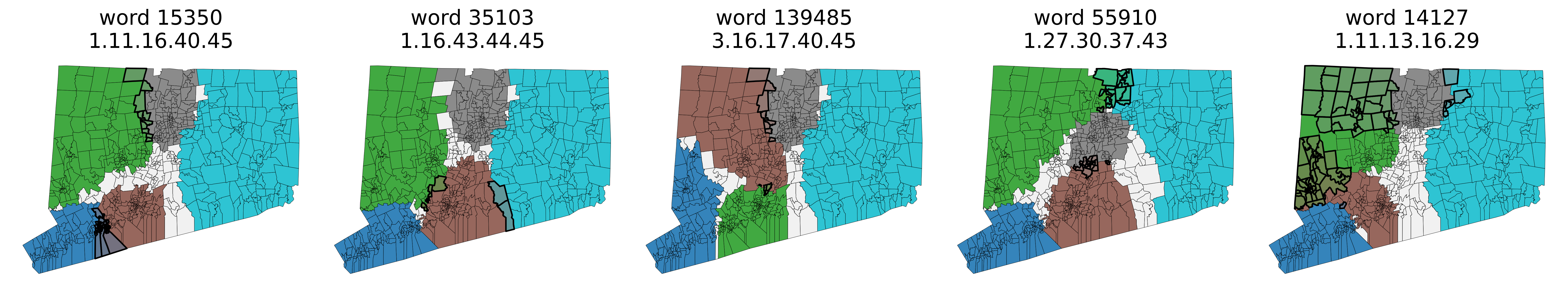}
    \caption{Similar to Figure \ref{fig:ct_top5_word_before_prune} but for the pruned set $\mathcal{S}_\eta$ produced by the two-stage greedy subcover. Pruning removes near-duplicate strata while maintaining coverage of the sampled ensemble, yielding a more compact and interpretable set of representative words. We again display the letter indices below the word, e.g. word 15350 is made up of letters 1, 11, 16, 40, and 45.}
    \label{fig:ct_top5_word_after_prune}
\end{figure}

To improve interpretability and reduce redundancy while preserving coverage of the observed plans, we construct the pruned set $\mathcal{S}_\eta$ using the two-stage greedy subcover in Algorithm~\ref{alg:eta_greedy_subcover}.
Here we take $\eta=1.0$, which selects $50$ representative words and covers all $49{,}961$ processed plans.
Figure~\ref{fig:ct_top5_word_after_prune} shows that, after pruning, the top words are more diverse and correspond to distinct large-scale plan motifs rather than minor variations within the same mode.
Word $15350$ has the largest coverage and appears in both Figure~\ref{fig:ct_top5_word_before_prune} and Figure~\ref{fig:ct_top5_word_after_prune}; it is selected first in the initial greedy subcover of Algorithm~\ref{alg:eta_greedy_subcover}.
This word is also close to the $2022$ enacted plan.
The southwestern letter (in blue) remains a strong district-level motif after pruning, while the other four districts show more variation across the retained words.

\begin{figure}[!htbp]
    \centering
    \includegraphics[width=1.0\linewidth]{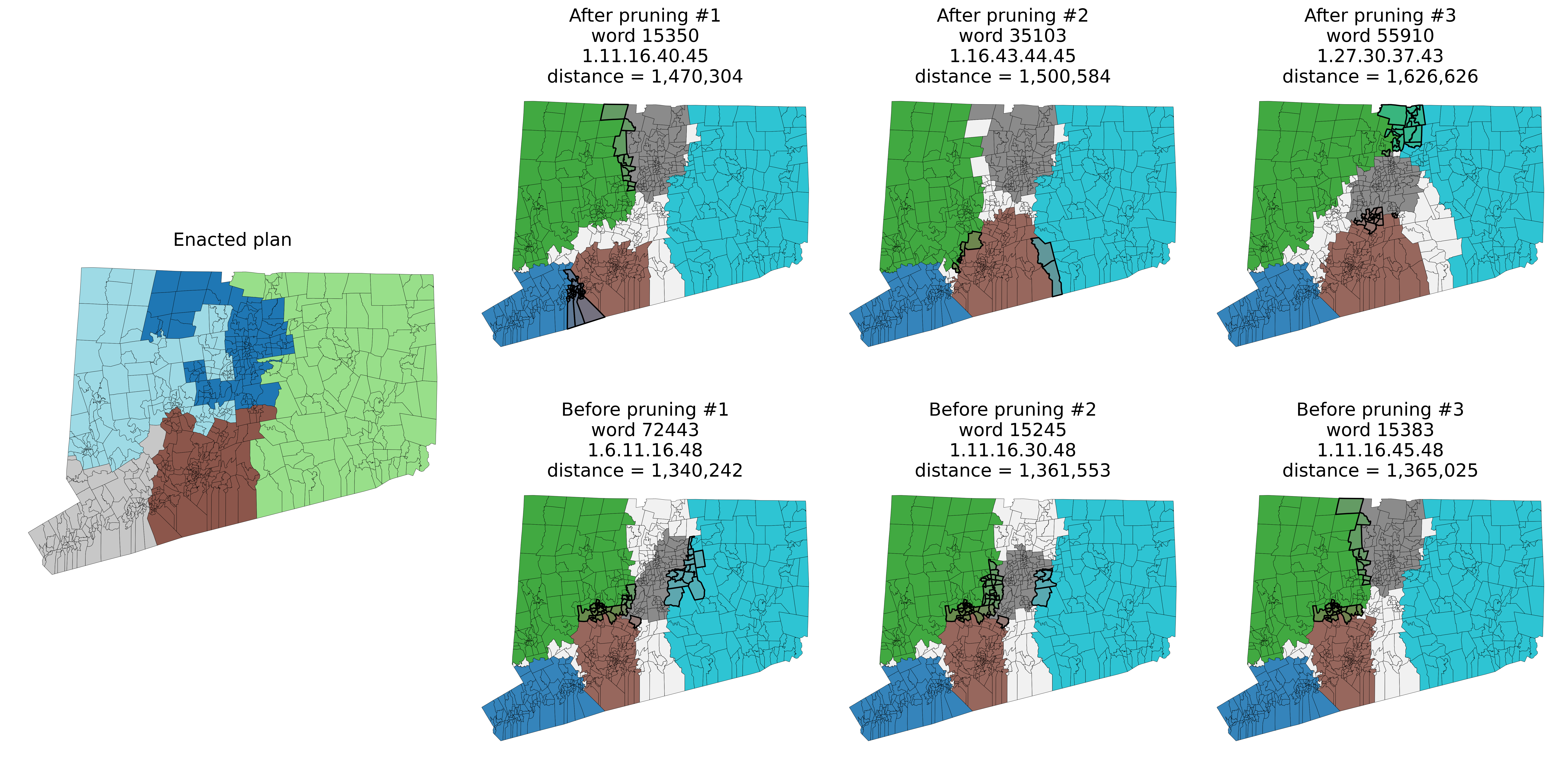}
    \caption{Closest words to the enacted plan. The first and second rows of the $2\times 3$ word plots show the three closest words after and before pruning, respectively. For each word displayed, we annotate its distance to the enacted plan.}
    \label{fig:ct_enacted_words}
\end{figure}

It is also interesting to see how the enacted plan fits into our strata, even though it follows a quite different measure.
In practice, redistricting plans are typically proposed by legislators and accepted through a political process.
See Figure \ref{fig:ct_enacted_words} for a visualization of the nearest words to the $2022$ enacted plan.
We compare the three closest words before pruning and in the two-stage pruned set $\mathcal{S}_\eta$.
As expected, the minimum distance between the enacted plan and the representative word set increases after pruning, from about $1.34$M to $1.47$M in the displayed distance.
Nevertheless, the closest word in the pruned word set successfully captures the macroscopic spatial structure of the plan.
Note that we do not see any highly specific local boundary geometry even before pruning, e.g. the dark blue district in the north-central region of the enacted plan.
These localized features are intentionally smoothed out during the hierarchical clustering phase, which abstracts districts into representative letters.

\begin{figure}[!htbp]
    \centering
    \includegraphics[width=1.0\linewidth]{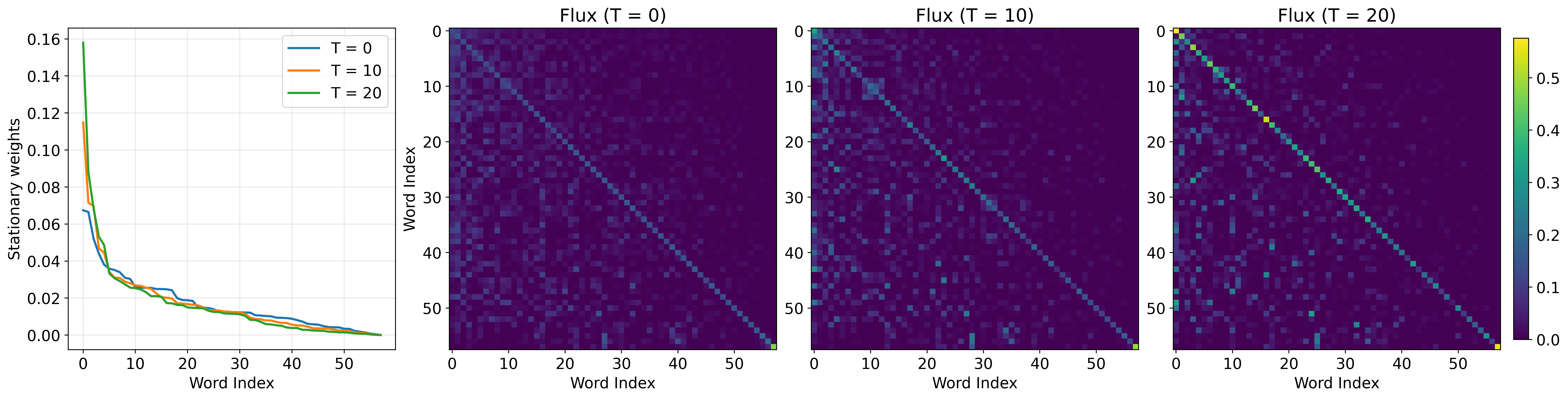}
    \caption{Stratum weights and flux matrices for $T=0,10,20$.}
    \label{fig:ct_stat_flux}
\end{figure}

\begin{figure}[!htbp]
    \centering
    \includegraphics[width=1.0\linewidth]{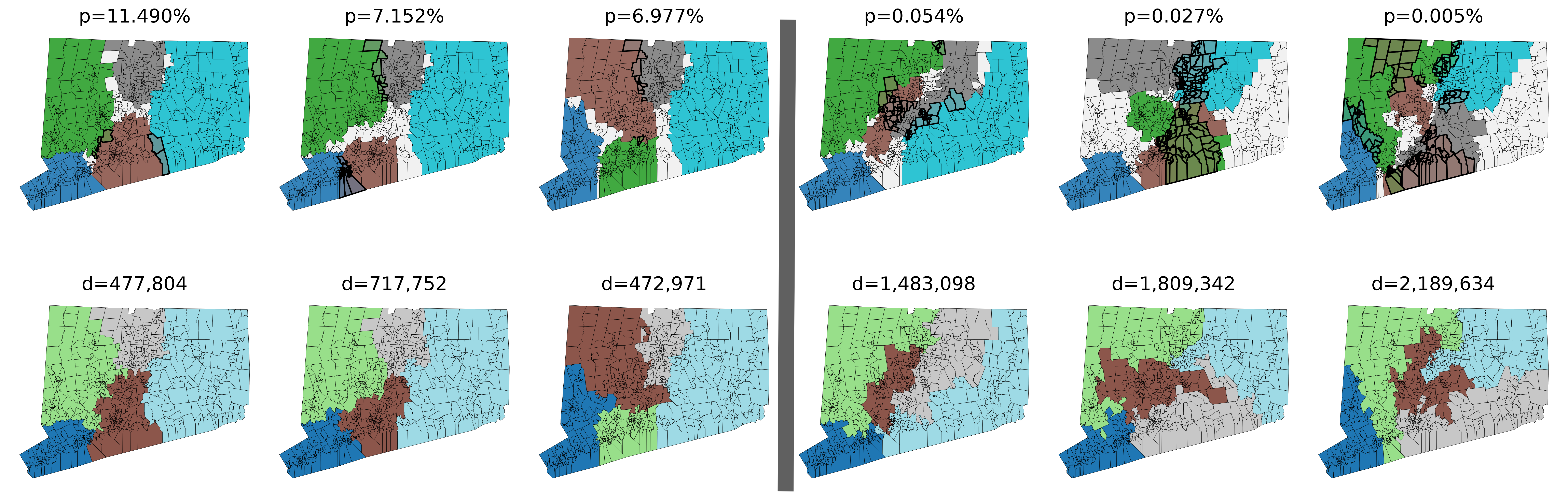}
    \caption{Top and bottom three words according to the stationary distribution at $T=10$. Words are shown in the first row, and example plans attached to those words are shown in the second row.}
    \label{fig:ct_top_bot_words}
\end{figure}

Next, we compute stationary distributions using multiple choices of temperatures while keeping the range mapping function $\kappa$ fixed.
Typically, we could tune $T$ and $\kappa$ (see \eqref{eq:psi_relative}) based on empirical observations from stratified sampling.
Since we do not implement stratified sampling in the current paper, we pick a generic $\kappa$ 
that normalizes distances to pruned neighbors by the total population of the state. 
We vary the temperature over $T=0,10,20$ in Figure \ref{fig:ct_stat_flux}.
Specifically, we include the case of $T=0$ (the counting measure) as an example that is $\kappa$ independent.
At the intermediate temperature $T=10$, we see reasonable overlaps across most strata except for the smallest one, which corresponds to an atypical plan.
See Figure \ref{fig:ct_top_bot_words} for more details, where we illustrate the three largest and three smallest strata together with example plans attached to those words.

\subsection{Stratified sampling on other measures}
In this section, we study how our strata construction generalizes to other measures.
As in~\cite{cyclewalk}, we sample from a family of unnormalized probability distributions proportional to $\exp\{-\gamma J_{\mathrm{tree}}(x)-c_\gamma J_{\mathrm{iso}}(x)\}$ parametrized by $(\gamma,c_\gamma)$, where $J_{\mathrm{tree}}$ is the log spanning-forest count and $J_{\mathrm{iso}}$ is the sum of district isoperimetric scores.
We refer to $c_\gamma$ as the iso parameter.
Intuitively, larger iso parameters tilt the target distribution toward smaller isoperimetric scores, i.e., more compact districts.
Meanwhile, increasing $\gamma$ from $0$ to $1$ tilts the distribution more toward uniform partitions, which are less compact.

\begin{figure}[!htbp]
    \centering
    \includegraphics[width=1.0\linewidth]{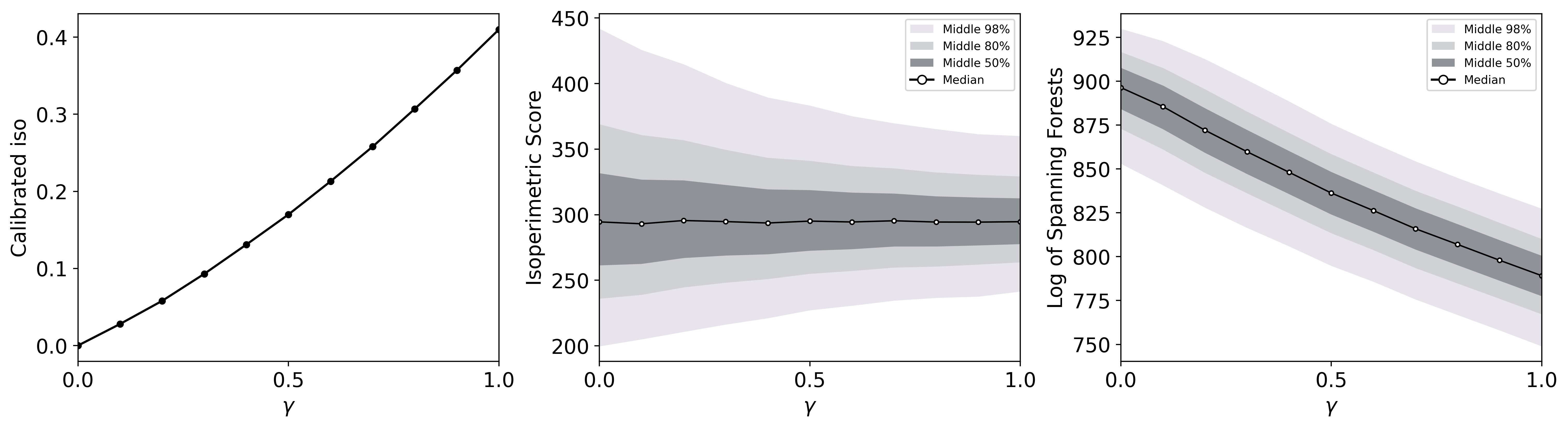}
    \caption{Calibration of the isoperimetric parameter for target measures indexed by $\gamma$. Left: calibrated iso value for each $\gamma$. Middle: isoperimetric-score distributions, with shaded bands showing the middle $50\%$, $80\%$, and $98\%$ intervals and markers showing medians. Right: distributions of log spanning-forest counts.}
    \label{fig:iso_calibration}
\end{figure}

\begin{figure}[!htbp]
    \centering
    \includegraphics[width=0.92\linewidth]{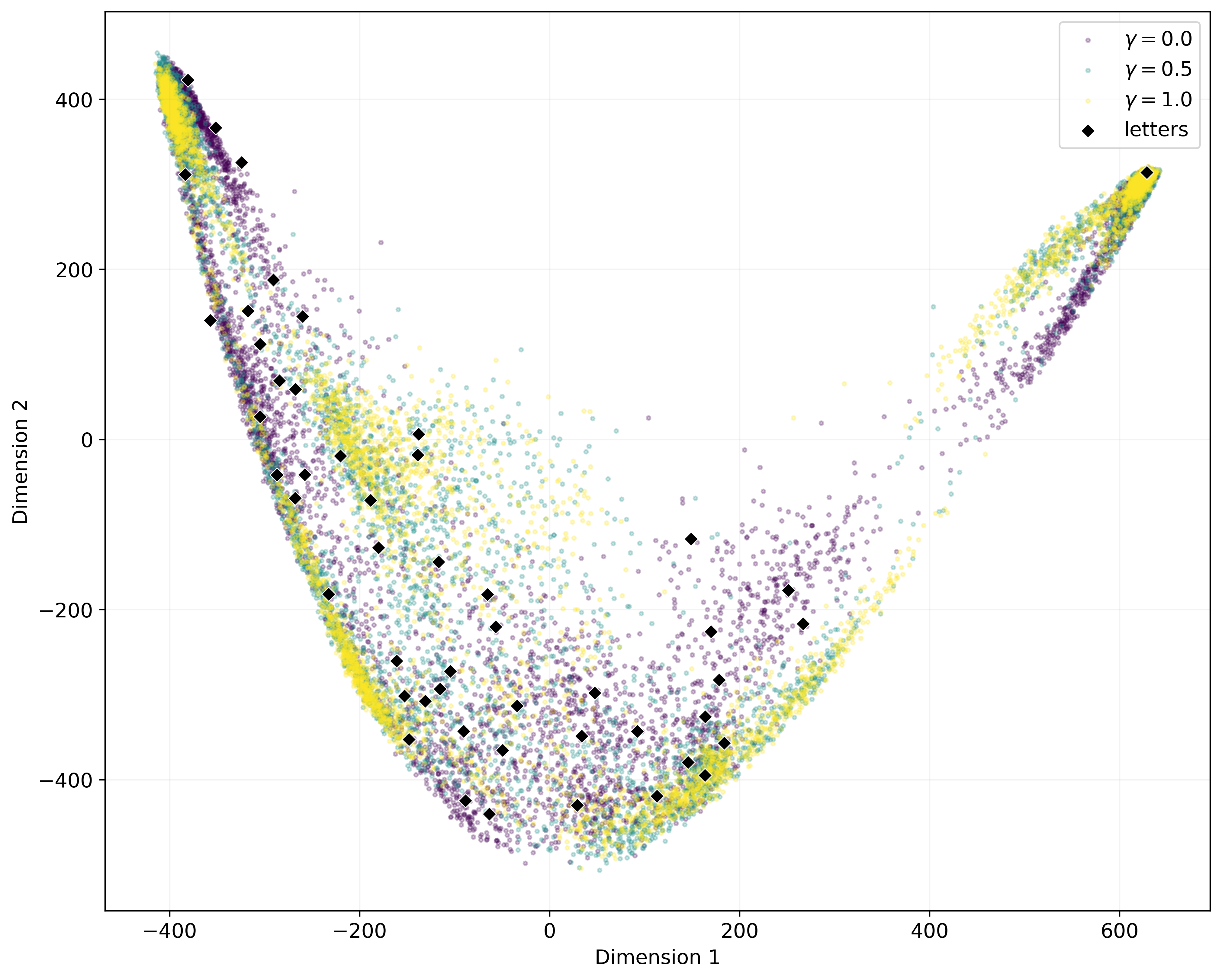}
    \caption{Classical multidimensional scaling embedding of Connecticut sampled districts for $\gamma\in\{0,0.5,1.0\}$ using seed A. Colors indicate $\gamma$, and black diamonds denote the representative letters learned from the baseline ensemble.}
    \label{fig:ct_cmds_overlay_seed_A}
\end{figure}

As a first step, we calibrate the iso parameter to obtain, for each $\gamma$ from $0$ to $1$, an iso parameter that produces samples with an isoperimetric ratio comparable to that of $(0,0)$.
We see that the relationship between $\gamma$ and the calibrated iso parameters is roughly linear, which is consistent with the observations in~\cite{cyclewalk} on North Carolina (see Figure~\ref{fig:iso_calibration}).
After calibration, the isoperimetric-score distributions remain comparable across $\gamma$, while the distribution of log spanning-forest counts shifts downward as $\gamma$ increases.
For each calibrated target measure, we generate two independent sets of CycleWalk samples of size 50,000 using two different random seeds $A$ and $B$.
We construct the alphabet and representative word set from run A at $\gamma=0$ and use them to analyze both runs for every value of $\gamma$.
Comparing the two runs helps determine whether our results are stable across independent samples.
Figure~\ref{fig:ct_cmds_overlay_seed_A} gives a geometric view of how the sampled district distribution changes as $\gamma$ varies.
To produce this visualization, we compute pairwise weighted $\ell^1$ distances between districts sampled from the three target distributions and project them to two dimensions using classical multidimensional scaling~\cite{torgerson1952multidimensional}.
The three point clouds occupy a common low-dimensional geometry, but their mass shifts as $\gamma$ changes.
This suggests that the alphabet computed with $\gamma=0$, shown as black diamonds in Figure \ref{fig:ct_cmds_overlay_seed_A}, remains informative for other target distributions in this family, while becoming less centered, though still with good coverage, on the target distribution as $\gamma$ moves away from $0$.

\begin{figure}[!htbp]
    \centering
    \includegraphics[width=1.0\linewidth]{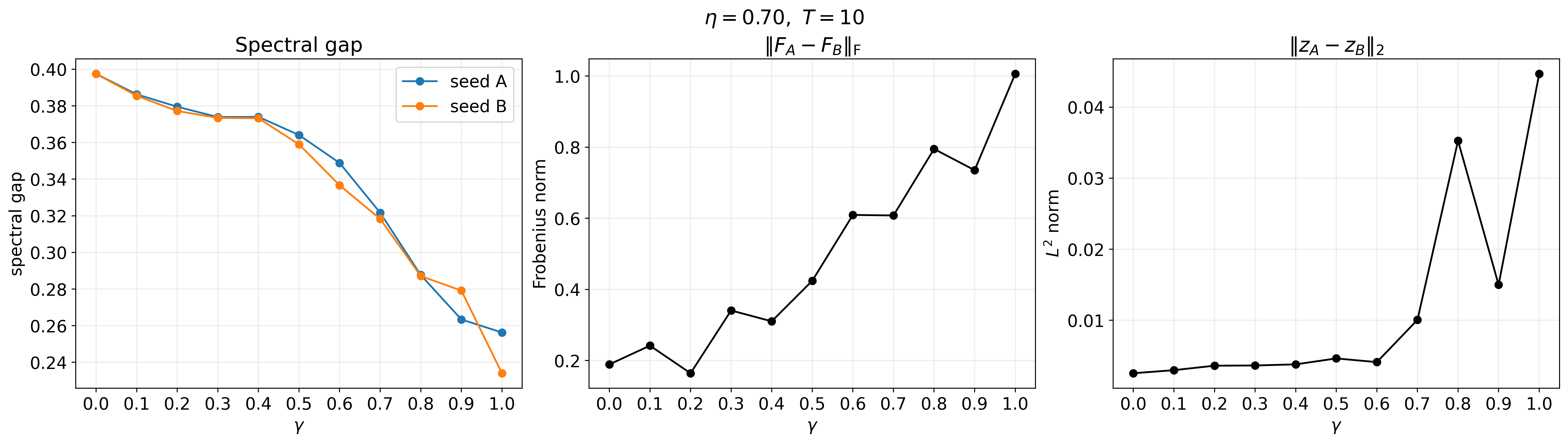}
    \caption{Using a fixed set of strata, taken at the spanning forest measure (i.e., $\gamma = 0$ and `easy' to sample), we show a comparison of two independent runs as $\gamma$ varies, with $\eta=0.70$ and $T=10$. Left: spectral gap of the flux matrix $F$ in \eqref{eq:flux}. Middle: Frobenius norm $\|F_A-F_B\|_F$. Right: $\ell^2$ distance $\|z_A-z_B\|_2$ between the invariant measures on strata. 
    }
    \label{fig:ct_gamma_eta_0.70_T_10}
\end{figure}

In practice, we would like to construct strata using samples from $\gamma=0$, where sampling is relatively cheap, and then use those strata to study other values of $\gamma$.
For this to be reasonable, the resulting words should still have enough overlap under the other target distributions.
As a first check, we fix $\eta=0.70$ and $T=10$. We compute and compare the fluxes and invariant measures across two independent runs (using one set of fixed strata from a single run at $\gamma=0$) to ensure that the strata taken from one run will provide consistent results across runs.
Figure~\ref{fig:ct_gamma_eta_0.70_T_10} shows that the qualitative spectral-gap behavior is consistent across seeds.
The differences between the two flux matrices and stationary distributions grow for larger $\gamma$, which may reflect slower convergence of the underlying chains as the target moves farther from the tree-count measure.
For other combinations of $\eta$ and $T$, we see patterns that are qualitatively similar.
\FloatBarrier

\begin{figure}[!htbp]
    \centering
    \includegraphics[width=1.0\linewidth]{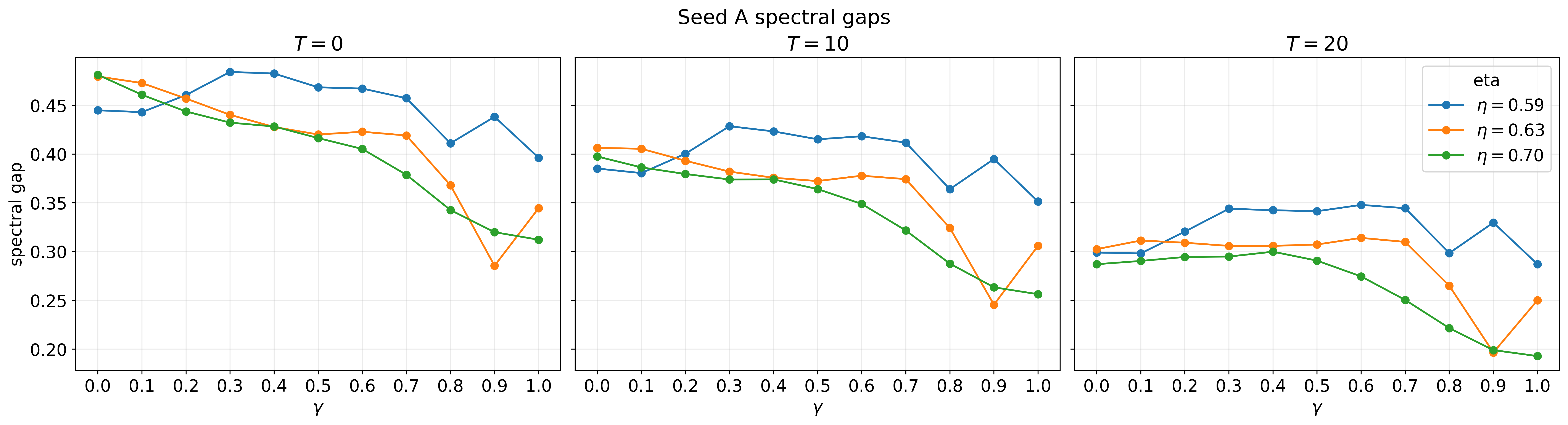}
    \caption{Spectral gap of the flux matrix $F$ in \eqref{eq:flux} for seed A as $\gamma$ varies. Each panel fixes the temperature $T$ in the POU kernel \eqref{eq:psi_relative}, while colors indicate the threshold parameter $\eta$ used in Algorithm~\ref{alg:eta_greedy_subcover}.}
    \label{fig:ct_gamma_spectral_gap_seed_A_by_temperature}
\end{figure}

We then examine how the POU temperature and pruning threshold affect this diagnostic.
We select threshold parameters $\eta=0.59,0.63,0.70$, which produce word sets of sizes $188,103,50$, respectively, and compute flux matrices for temperatures $T=0,10,20$.
Together, these choices give nine experimental configurations.
Figure~\ref{fig:ct_gamma_spectral_gap_seed_A_by_temperature} shows the spectral gap of the resulting flux matrices as $\gamma$ varies.
The spectral gap decreases for two related reasons: larger $\gamma$ moves the target distribution away from the $\gamma=0$ distribution used to construct the strata, while larger $T$ concentrates each word on closer plans and therefore reduces its effective support.

\section{Future Directions}\label{sec:chap4 future_directions}

The natural next step is to use this stratification to run stratified sampling to estimate a plan-level observable of interest $f$, for instance, Democratic vote shares.
Let $\pi(x)\propto e^{-J(x)}$ be the target distribution and let $\{\phi_s\}_{s\in\mathcal{S}}$ be the POU constructed by our pipeline.
Recall that the corresponding POU strata and their weights are
\begin{align}
    \pi_s(x):=\frac{\phi_s(x)\pi(x)}{z_s},
    \qquad
    z_s:=\mathbb{E}_{\pi}[\phi_s].
\end{align}
The POU property therefore gives the exact decomposition
\begin{align}
\label{eq:stratifiedobs}
    \mathbb{E}_\pi[f]=\sum_{s\in\mathcal{S}}z_s\mathbb{E}_{\pi_s}[f].
\end{align}

For future sampling, we propose a distinct family of auxiliary biased distributions
\begin{align}
    q_s(x):=\frac{e^{\phi_s(x)}\pi(x)}{Z_s},
    \qquad
    Z_s:=\mathbb{E}_{\pi}[e^{\phi_s}],
\end{align}
or, equivalently,
\begin{align}
    q_s(x)\propto e^{-J_s^{\mathrm{bias}}(x)},
    \qquad
    J_s^{\mathrm{bias}}(x):=J(x)-\phi_s(x).
\end{align}
These auxiliary distributions are different from the POU strata $\pi_s$ and are introduced only as a sampling mechanism.
Using the same method (e.g., Cycle Walk), we would sample one chain for each $q_s$, adjusting the Metropolis--Hastings steps to satisfy detailed balance with respect to $q_s$.
Standard fixed-point iteration methods can recover the relative normalizing factors $Z_s$.
Expectations under the POU strata can then be recovered from samples of the auxiliary distributions using
\begin{align}
    \mathbb{E}_{\pi_s}[f]
    =\frac{\mathbb{E}_{q_s}\!\left[f\phi_s e^{-\phi_s}\right]}
    {\mathbb{E}_{q_s}\!\left[\phi_s e^{-\phi_s}\right]},
    \qquad
    z_s=Z_s\mathbb{E}_{q_s}\!\left[\phi_s e^{-\phi_s}\right],
\end{align}
where the resulting $z_s$ are normalized so that $\sum_{s\in\mathcal{S}}z_s=1$ when the $Z_s$ are known only up to a common factor.
Ideally, each auxiliary chain focuses on plans around a different word $s\in\mathcal{S}$, which allows for more accurate estimates of the observables.

Although we do not yet sample the auxiliary distributions $q_s$, we perform an offline reconstruction diagnostic using the existing chains and the POU identity \eqref{eq:stratifiedobs}.
We consider the order statistics of the partisan leans under the 2020 Presidential votes. To compute these marginal distributions, we take each district within a plan and compute the number of Democratic votes divided by the number of Democratic and Republican votes to compute the Democratic partisan lean. We then order the districts from least to most to sample an empirical distribution on the ordered marginals (i.e., we obtain an empirical marginal distribution on the most Republican, second-most Republican, etc. districts).

We examine these distributions on two independent chains (seed A and seed B) which begin with different initial conditions. We then construct our strata using seed A and different length scales $\eta$ for the greedy set cover. Let $\{x_B^{(m)}\}_{m=1}^M$ denote the samples from seed B. From these samples, we compute the observed fluxes and empirical POU weights
\begin{align}
    \widehat{z}_s:=\frac{1}{M}\sum_{m=1}^M\phi_s(x_B^{(m)}),
\end{align}
and estimate the observable within each POU stratum as
\begin{align}
    \widehat{\mathbb{E}}_{\pi_s}[f]
    :=\frac{\displaystyle\sum_{m=1}^M \phi_s(x_B^{(m)}) f(x_B^{(m)})}
    {\displaystyle\sum_{m=1}^M \phi_s(x_B^{(m)})}.
\end{align}
We substitute these quantities into \eqref{eq:stratifiedobs} for a family of observables $f$ to reconstruct and estimate the marginal densities.
For each district rank, we measure the probability that its Democratic vote share falls within some specific 0.2-percentage-point bin (e.g. 0-0.2\%, 0.2-0.4\%...), which is then used to produce a histogram over the vote share for each rank-ordered district (i.e., the most, 2nd most,... and least Republican district).

\begin{figure}[!htbp]
    \centering
    \includegraphics[width=1.0\linewidth]{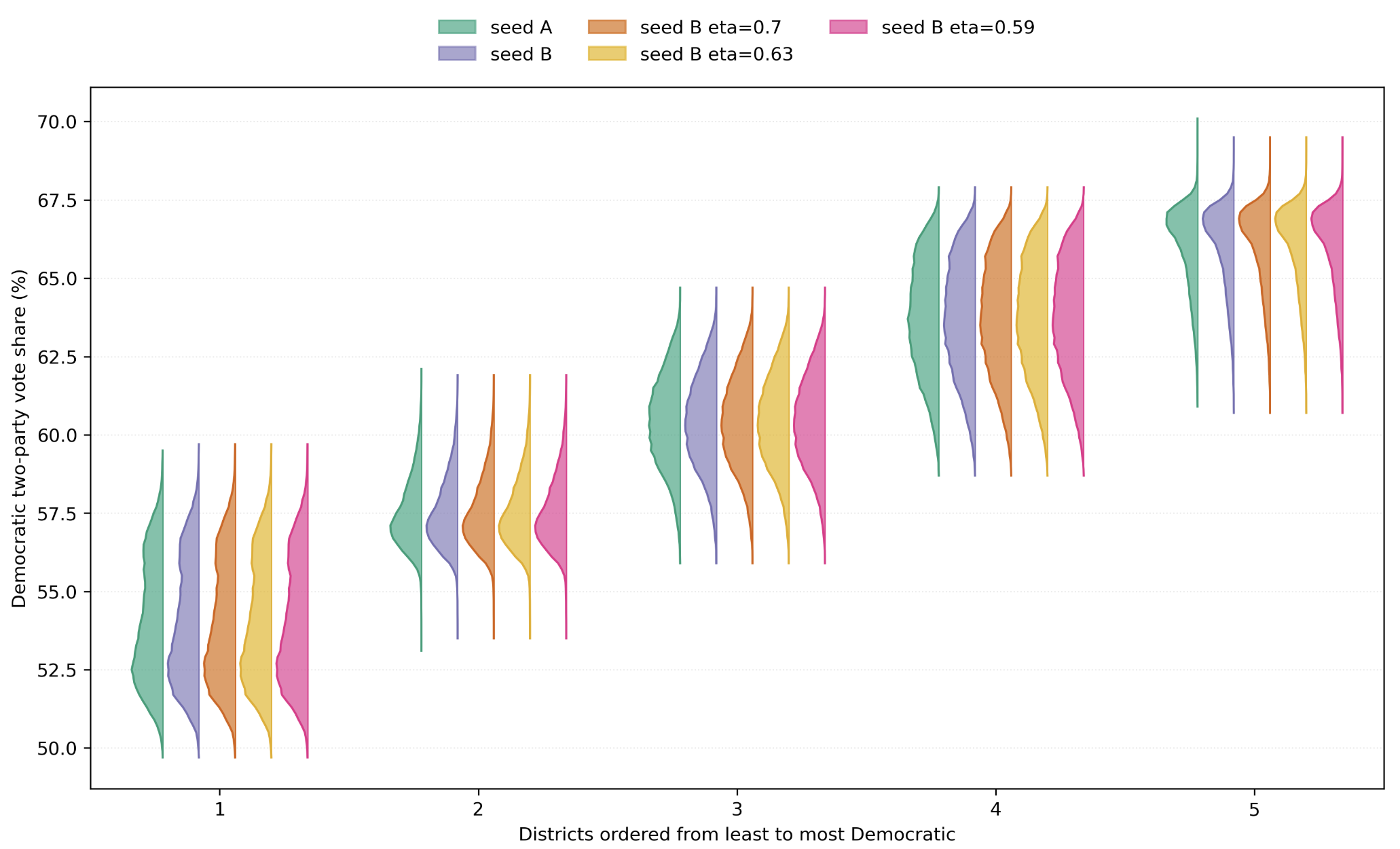}
    \caption{We compare rank-ordered marginal densities of the partisan vote share of independent runs (seeds A and B), along with reconstructions using \eqref{eq:stratifiedobs} with strata from seed A on samples from seed B for strata of various sizes (parametrized by $\eta$).}
    \label{fig:observablereconstruction}
\end{figure}

We plot the results in Figure~\ref{fig:observablereconstruction}, showing the marginal densities for the two independent seeds and three reconstructions using strata defined and pruned from seed A on samples obtained from seed B. In all cases, we see excellent agreement.

Another interesting direction is to adaptively update the alphabet as we obtain new data, so that the alphabet still captures the motifs in district space.
This can be useful when the original sample used to construct the alphabet is too small, or when the target distribution changes drastically, causing new motifs to appear.
Re-running hierarchical clustering from scratch after each new sample can be costly.
On the other hand, most hierarchical procedures are not designed for streaming data.
Online hierarchical clustering has been studied, but under different objectives~\cite{menon2019online}.
If we expect localized updates, i.e., $K$ does not change, we can assign new data to the closest centroids (i.e., Rocchio classification) and run a few $K$-median iterations (if necessary) to ensure that the letters reflect their corresponding clusters.
While this approach is intuitive and computationally cheap, the hierarchical structure is broken and it is not clear whether the theoretical guarantees~\cite{DBLP:journals/corr/abs-2409-01010} are preserved.

Finally, sampling parameters can substantially change district distributions.
For a target such as $\gamma=1.0$, we may learn better letters from an intermediate distribution, such as $\gamma=0.5$, than from a distant baseline such as $\gamma=0.0$.
We can also form a shared alphabet by taking, and possibly pruning, the union of letters learned across several values of $\gamma$.

\section{Conclusion}\label{sec:chap4 conclusion}

In this paper, we introduce a computational pipeline for constructing and diagnosing candidate strata on the space of redistricting plans.
We first aggregate districts across sampled redistricting plans and apply a hierarchical correlation clustering algorithm \hccultra to identify representative districts (``letters'').
We then use a beam search to find a small set of ``words'' around each plan.
Each word is an unordered tuple of letters and can be viewed as a representative plan.
We take the union of these words to form the strata for stratified sampling via a partition of unity that softly assigns plans to words based on their distances.
To reduce redundancy among words while maintaining coverage, we apply a two-stage greedy set-cover approximation.
We illustrate our pipeline using samples generated with real-world data based on the state of Connecticut.
The present results assess the representation, empirical coverage, and overlap of the resulting strata, but do not test a complete stratified sampler.
Implementing that sampler and determining whether it improves sampling efficiency or reduces estimator variance remain future work.

\newpage
\bibliographystyle{plain}
\bibliography{sample}

@article{torgerson1952multidimensional,
  title={Multidimensional scaling: I. Theory and method},
  author={Torgerson, Warren S},
  journal={Psychometrika},
  volume={17},
  number={4},
  pages={401--419},
  year={1952},
  publisher={Springer-Verlag}
}

@article{fifield2015new,
  title={A new automated redistricting simulator using markov chain monte carlo},
  author={Fifield, Benjamin and Higgins, Michael and Imai, Kosuke and Tarr, Alexander},
  journal={Work. Pap., Princeton Univ., Princeton, NJ},
  year={2014}
}

@article{mcdonald2008united,
  title={United States election project},
  author={McDonald, Michael},
  journal={United States Elections Project},
  year={2008}
}

@article{mattingly2019expert,
  title={Expert Report on the North Carolina State Legislature},
  author={Mattingly, Jonathan},
  journal={Lewis v. Common Cause},
  year={2019}
}

@misc{mattingly2021expert,
  title={Expert report on the North Carolina State Legislature and Congressional redistricting (corrected version)},
  author={Mattingly, Jonathan C},
  year={2021},
  publisher={Harper v. Hall, North Carolina Superior Court, Wake County}
}

@article{kastner2011umbrella,
  title={Umbrella sampling},
  author={K{\"a}stner, Johannes},
  journal={Wiley Interdisciplinary Reviews: Computational Molecular Science},
  volume={1},
  number={6},
  pages={932--942},
  year={2011},
  publisher={Wiley Online Library}
}

@article{kumar1992weighted,
  title={The weighted histogram analysis method for free-energy calculations on biomolecules. I. The method},
  author={Kumar, Shankar and Rosenberg, John M and Bouzida, Djamal and Swendsen, Robert H and Kollman, Peter A},
  journal={Journal of computational chemistry},
  volume={13},
  number={8},
  pages={1011--1021},
  year={1992},
  publisher={Wiley Online Library}
}

@article{torrie1977nonphysical,
  title={Nonphysical sampling distributions in Monte Carlo free-energy estimation: Umbrella sampling},
  author={Torrie, Glenn M and Valleau, John P},
  journal={Journal of computational physics},
  volume={23},
  number={2},
  pages={187--199},
  year={1977},
  publisher={Elsevier}
}

@article{sokal1958statistical,
  title={A statistical method for evaluating systematic relationships},
  author={Sokal, Robert R. and Michener, Charles D.},
  journal={The University of Kansas Science Bulletin},
  volume={38},
  number={22},
  pages={1409--1438},
  year={1958},
  url={https://www.biodiversitylibrary.org/part/385476}
}

@article{defays1977efficient,
  title={An efficient algorithm for a complete link method},
  author={Defays, Daniel},
  journal={The computer journal},
  volume={20},
  number={4},
  pages={364--366},
  year={1977},
  publisher={Oxford University Press}
}

@article{sibson1973slink,
  title={SLINK: an optimally efficient algorithm for the single-link cluster method},
  author={Sibson, Robin},
  journal={The computer journal},
  volume={16},
  number={1},
  pages={30--34},
  year={1973},
  publisher={Oxford University Press}
}

@article{murtagh2012algorithms,
  title={Algorithms for hierarchical clustering: an overview},
  author={Murtagh, Fionn and Contreras, Pedro},
  journal={Wiley interdisciplinary reviews: data mining and knowledge discovery},
  volume={2},
  number={1},
  pages={86--97},
  year={2012},
  publisher={Wiley Online Library}
}

@article{deford2021recombination,
  title={Recombination: A family of Markov chains for redistricting},
  author={DeFord, Daryl and Duchin, Moon and Solomon, Justin},
  journal={Harvard Data Science Review},
  volume={3},
  number={1},
  pages={3},
  year={2021},
  publisher={The MIT Press}
}

@article{menon2019online,
  title={Online hierarchical clustering approximations},
  author={Menon, Aditya Krishna and Rajagopalan, Anand and Sumengen, Baris and Citovsky, Gui and Cao, Qin and Kumar, Sanjiv},
  journal={arXiv preprint arXiv:1909.09667},
  year={2019}
}

@book{eckman2019apportionment,
  title={Apportionment and redistricting process for the US House of Representatives},
  author={Eckman, Sarah J},
  year={2019},
  publisher={Congressional Research Service}
}

@article{deford2019redistricting,
  title={Redistricting reform in Virginia: Districting criteria in context},
  author={DeFord, Daryl and Duchin, Moon},
  journal={Virginia Policy Review},
  volume={12},
  number={2},
  pages={120--146},
  year={2019}
}

@article{dinner2020stratification,
  title={Stratification as a general variance reduction method for Markov chain Monte Carlo},
  author={Dinner, Aaron R and Thiede, Erik H and Koten, Brian Van and Weare, Jonathan},
  journal={SIAM/ASA journal on uncertainty quantification},
  volume={8},
  number={3},
  pages={1139--1188},
  year={2020},
  publisher={SIAM}
}

@article{shirts2008statistically,
  title={Statistically optimal analysis of samples from multiple equilibrium states},
  author={Shirts, Michael R and Chodera, John D},
  journal={The Journal of chemical physics},
  volume={129},
  number={12},
  year={2008},
  publisher={AIP Publishing}
}

@article{matthews2018umbrella,
  title={Umbrella sampling: A powerful method to sample tails of distributions},
  author={Matthews, Charles and Weare, Jonathan and Kravtsov, Andrey and Jennings, Elise},
  journal={Monthly Notices of the Royal Astronomical Society},
  volume={480},
  number={3},
  pages={4069--4079},
  year={2018},
  publisher={Oxford University Press}
}

@article{dinner2018trajectory,
  title={Trajectory stratification of stochastic dynamics},
  author={Dinner, Aaron R and Mattingly, Jonathan C and Tempkin, Jeremy OB and Koten, Brian Van and Weare, Jonathan},
  journal={Siam Review},
  volume={60},
  number={4},
  pages={909--938},
  year={2018},
  publisher={SIAM}
}

@article{DBLP:journals/corr/abs-2409-01010,
  author       = {Joon{-}Hyeok Yim and
                  Anna C. Gilbert},
  title        = {Fitting trees to {$\ell_1$}-hyperbolic distances},
  journal      = {CoRR},
  volume       = {abs/2409.01010},
  year         = {2024},
  url          = {https://doi.org/10.48550/arXiv.2409.01010},
  doi          = {10.48550/ARXIV.2409.01010},
  eprinttype    = {arXiv},
  eprint       = {2409.01010},
  bibsource    = {dblp computer science bibliography, https://dblp.org}
}

@article{autry2020multi,
  title={Multi-scale merge-split Markov chain Monte Carlo for redistricting},
  author={Autry, Eric A and Carter, Daniel and Herschlag, Gregory and Hunter, Zach and Mattingly, Jonathan C},
  journal={arXiv preprint arXiv:2008.08054},
  year={2020}
}

@article{autry2023metropolized,
  title={Metropolized forest recombination for Monte Carlo sampling of graph partitions},
  author={Autry, Eric and Carter, Daniel and Herschlag, Gregory J and Hunter, Zach and Mattingly, Jonathan C},
  journal={SIAM Journal on Applied Mathematics},
  volume={83},
  number={4},
  pages={1366--1391},
  year={2023},
  publisher={SIAM}
}

@article{cyclewalk,
  author       = {Daryl R. DeFord and
                  Gregory Herschlag and
                  Jonathan C. Mattingly},
  title        = {A Cycle Walk for Sampling Measures on Spanning Forests for Redistricting},
  journal      = {CoRR},
  volume       = {abs/2509.08629},
  year         = {2025},
  url          = {https://doi.org/10.48550/arXiv.2509.08629},
  doi          = {10.48550/ARXIV.2509.08629},
  eprinttype    = {arXiv},
  eprint       = {2509.08629},
  bibsource    = {dblp computer science bibliography, https://dblp.org}
}

@article{mattingly2014redistricting,
  title={Redistricting and the Will of the People},
  author={Mattingly, Jonathan C and Vaughn, Christy},
  journal={arXiv preprint arXiv:1410.8796},
  year={2014}
}

@misc{redistrictingdatahub,
  title        = {Redistricting Data Hub},
  howpublished = {\url{http://redistrictingdatahub.org}},
  note         = {Accessed: 2023-09-30},
  key          = {Redistricting Data Hub},
}

@misc{LewisVCommonCause,
    key = {Common Cause v. Lewis},
    title = {{Common Cause v. Lewis}},
    howpublished = {No. 18 CVS 014001 (N.C. Super. Ct., Wake County)},
    month = {September~3},
    year = {2019},
    url = {https://www.nccourts.gov/assets/inline-files/18-CVS-14001_Final-Judgment.pdf}
}

@misc{HarperVHallMoore,
    key = {Harper v. Hall},
    title = {{Harper v. Hall}},
    howpublished = {380 N.C. 317, 868 S.E.2d 499},
    year = {2022},
    url = {https://law.justia.com/cases/north-carolina/supreme-court/2022/413pa21.html}
}

@misc{LWVvPA,
    key = {League of Women Voters of Pennsylvania v. Commonwealth},
    title = {{League of Women Voters of Pennsylvania v. Commonwealth}},
    howpublished = {178 A.3d 737 (Pa.)},
    year = {2018},
    url = {https://www.pacourts.us/Storage/media/pdfs/20211215/172052-feb.7,2018-majorityopinion.pdf}
}

@misc{RuchoVCC,
    key = {Rucho v. Common Cause},
    title = {{Rucho v. Common Cause}},
    howpublished = {588 U.S. 684},
    year = {2019},
    url = {https://www.supremecourt.gov/opinions/18pdf/18-422_9ol1.pdf}
}

@misc{CovingtonvNC,
    key = {Covington v. North Carolina},
    title = {{Covington v. North Carolina}},
    howpublished = {316 F.R.D. 117 (M.D.N.C.)},
    year = {2016},
    url = {https://openjurist.org/316/frd/117/covington-v-north-carolina}
}

@misc{duchinPAreport,
  title={Outlier analysis for {Pennsylvania} congressional redistricting},
  author={Duchin, Moon},
  howpublished={\href{https://mggg.org/uploads/md-report.pdf}{Report}},
  month=feb,
  year={2018},
  url={https://mggg.org/uploads/md-report.pdf}
}

@misc{AllenVMilligan,
    key = {Allen v. Milligan},
    title = {{Allen v. Milligan}},
    howpublished = {599 U.S. 1},
    year = {2023},
    url = {https://www.supremecourt.gov/opinions/22pdf/21-1086_1co6.pdf}
}

@article{herschlag2020quantifying,
  title={Quantifying gerrymandering in north carolina},
  author={Herschlag, Gregory and Kang, Han Sung and Luo, Justin and Graves, Christy Vaughn and Bangia, Sachet and Ravier, Robert and Mattingly, Jonathan C},
  journal={Statistics and Public Policy},
  volume={7},
  number={1},
  pages={30--38},
  year={2020},
  publisher={Taylor \& Francis}
}

@article{Chen15,
  Author = {Chen, Jowei and Jonathan Rodden},
  Journal = {Election Law Journal},
  Number = {4},
  Pages = {331--345},
  Title = {Cutting through the Thicket: Redistricting Simulations and the Detection of Partisan Gerrymanders},
  Volume = {14},
  Year = {2015}}

@inproceedings{gurnee2021fairmandering,
  title={Fairmandering: A column generation heuristic for fairness-optimized political districting},
  author={Gurnee, Wes and Shmoys, David B},
  booktitle={SIAM Conference on Applied and Computational Discrete Algorithms (ACDA21)},
  pages={88--99},
  year={2021},
  organization={SIAM}
}

@article{mccartan2023sequential,
  title={Sequential Monte Carlo for sampling balanced and compact redistricting plans},
  author={McCartan, Cory and Imai, Kosuke},
  journal={The Annals of Applied Statistics},
  volume={17},
  number={4},
  pages={3300--3323},
  year={2023},
  publisher={Institute of Mathematical Statistics}
}

@article{cannon2026spanning,
  title={Spanning trees and redistricting: New methods for sampling and validation},
  author={Cannon, Sarah and Duchin, Moon and Randall, Dana and Rule, Parker},
  journal={SIAM Review},
  volume={68},
  number={2},
  pages={349--381},
  year={2026},
  publisher={SIAM}
}

@article{mcwhorter2025marked,
  title={The Marked Edge Walk: A Novel MCMC Algorithm for Sampling of Graph Partitions},
  author={McWhorter, Atticus and DeFord, Daryl},
  journal={arXiv preprint arXiv:2510.17714},
  year={2025}
}

@article{akitaya2026balanced,
  title={The Balanced Up-Down Walk},
  author={Akitaya, Hugo A and Cannon, Sarah and Herschlag, Gregory and Schoenbach, Gabe and Tapp, Kristopher and Tucker-Foltz, Jamie},
  journal={arXiv preprint arXiv:2602.11993},
  year={2026}
}

@article{chuang2025multiscale,
  title={Multiscale parallel tempering for fast sampling on redistricting plans},
  author={Chuang, Gabriel and Herschlag, Gregory and Mattingly, Jonathan},
  journal={Multiscale Modeling \& Simulation},
  volume={23},
  number={4},
  pages={1515--1550},
  year={2025},
  publisher={SIAM}
}

@article{o2026generalized,
  title={Generalized Sequential Monte Carlo Sampling for Redistricting Simulation},
  author={O'Sullivan, Philip and Imai, Kosuke and McCartan, Cory},
  journal={arXiv preprint arXiv:2603.22188},
  year={2026}
}

@article{fifield2020essential,
  title={The essential role of empirical validation in legislative redistricting simulation},
  author={Fifield, Benjamin and Imai, Kosuke and Kawahara, Jun and Kenny, Christopher T},
  journal={Statistics and Public Policy},
  volume={7},
  number={1},
  pages={52--68},
  year={2020},
  publisher={Taylor \& Francis}
}

\end{document}